\documentclass[a4paper,twocolumn,10pt,unpublished]{quantumarticle}
\pdfoutput=1

\usepackage[utf8]{inputenc}
\usepackage[english]{babel}
\usepackage[T1]{fontenc}

\usepackage{tikz}
\usepackage{lipsum}
\usepackage[numbers,sort&compress]{natbib}
\usepackage{listings}

\usepackage{hyperref}

\usepackage{graphicx}
\usepackage{dcolumn}
\usepackage{bm}
\usepackage{braket}
\usepackage{amsmath}
\usepackage{amssymb}
\usepackage{feynmp}
\usepackage{graphics}
\usepackage{color}

\usepackage{physics}
\usepackage{dsfont}

\usepackage[linesnumbered,ruled]{algorithm2e}

\usetikzlibrary{quantikz}

\definecolor{atomictangerine}{rgb}{1.0, 0.6, 0.4}
\definecolor{brightturquoise}{rgb}{0.03, 0.91, 0.87}
\definecolor{corn}{rgb}{0.98, 0.93, 0.36}

\definecolor{atomictangerine}{rgb}{1.0, 0.6, 0.4}
\definecolor{brightturquoise}{rgb}{0.03, 0.91, 0.87}
\definecolor{corn}{rgb}{0.98, 0.93, 0.36}
\definecolor{darkgoldenrod}{rgb}{0.72, 0.53, 0.04}
\definecolor{fluorescentorange}{rgb}{1.0, 0.75, 0.0}

\definecolor{capri}{rgb}{0.0, 0.75, 1.0}
\definecolor{caribbeangreen}{rgb}{0.0, 0.8, 0.6}

\usepackage{subcaption}

\begin{document}

\title{Resource-efficient Variational Compilation of Block-Encodings}

\author{Leon Rullkötter}
\affiliation{Fraunhofer IAO, Nobelstraße 12, 70569 Stuttgart, Germany}
\affiliation{Institute for Theoretical Physics III and Center for Integrated Quantum Science and Technology, Universität Stuttgart, Pfaffenwaldring 57, 70569 Stuttgart, Germany}
\email{leon.rullkoetter@iao.fraunhofer.de}
\author{Sebastian Weber}
\affiliation{Institute for Theoretical Physics III and Center for Integrated Quantum Science and Technology, Universität Stuttgart, Pfaffenwaldring 57, 70569 Stuttgart, Germany}
\author{Vamshi Mohan Katukuri}
\affiliation{Fraunhofer IAO, Nobelstraße 12, 70569 Stuttgart, Germany}
\author{Christian Tutschku}
\affiliation{Fraunhofer IAO, Nobelstraße 12, 70569 Stuttgart, Germany}
\author{Bharadwaj Chowdary Mummaneni}
\affiliation{Fraunhofer IAO, Nobelstraße 12, 70569 Stuttgart, Germany}
\maketitle

\begin{abstract}
    Block-encoding operators are one of the essential components in quantum algorithms based on Quantum Signal Processing. Their gate complexity largely determines the overall gate complexity of the full algorithm. Using variational methods, we compile single-ancilla block-encoding unitaries with near-optimal resource requirements for a large range of input matrices. We find that the number of variational parameters in the parameterized quantum circuit approaches the number of free parameters in the input matrices, depending on whether they are real, complex and/or hermitian. Additionally, symmetries present in the input matrix can be incorporated into the ansatz circuit, reducing the parameter count further and enhancing optimizability. While performing a variational compilation of block-encodings ceases to be computationally feasible for large system sizes, the constructed operators can be used as components of larger block-encodings via a linear combination of block-encodings.
\end{abstract}

\section{Introduction}
\label{section1}

Quantum Signal Processing (QSP)-type quantum algorithms have become the powerhouse of algorithmic developments in quantum computing. Starting with the introduction of QSP together with Qubitization to create the most efficiently scalable Hamiltonian Simulation~\cite{Low2016OptimalProcessing,Low2016HamiltonianQubitization}, subsequent efforts introduced further generalizations~\cite{Gilyen2018QuantumArithmetics,Motlagh2024} and were even able to unify a large number of existing quantum algorithms~\cite{Martyn2021AAlgorithms}. Central to these methods is a polynomial transformation of an input matrix $A$. While quantum computers can only perform unitary operations natively, $A$ is not necessarily unitary. A new form of data encoding was introduced termed block-encoding where a non-unitary input is encoded into a sub-block of a larger unitary $U_\mathrm{A}$ executable in a quantum circuit. This gives the freedom to make general $A \in \mathbb{C}^{m \times n}$ accessible to a quantum algorithm.

Designing efficient block-encoding operators is one of the key challenges. Established methods are the Linear Combinations of Unitaries (LCU)~\cite{Childs2012HamiltonianOperations} and oracle query models~\cite{Low2016HamiltonianQubitization}. While being able to generate encodings that are proven to give a quantum advantage in combination with QSP, they have considerable shortcomings when viewed in the light of near-term quantum computers with limited quantum resources. LCU demands a large number of multi-controlled gates and explicit oracle-based approaches often have high ancilla overheads. Both can suffer from an unfavorable scaling of a sub-normalization factor $\alpha$ reducing the amplitude of the input matrix and also do not scale well when transitioning to dense and unstructured matrices~\cite{Low2016HamiltonianQubitization}.

To bridge this gap and enable QSP applications on resource-limited devices, Kikuchi et al.~\cite{Kikuchi2023RealizationComputer} proposed to implement the block-encoding operator by variationally optimizing a hardware-efficient parameterized quantum circuit (PQC). Initial results demonstrated the ability to reduce circuit depths and width significantly. This approach, which we term Variational Compilation of Block-Encodings (VCBE), leverages the immense expressibility of PQCs and promises a path towards more resource-efficient block-encoding circuits for a wide range of target matrices.

While~\cite{Kikuchi2023RealizationComputer} demonstrated VCBE's potential, the minimal resources required for encoding a target exactly remained unclear. We demonstrate that VCBE ansatze can achieve an exact encoding using only a single ancilla qubit. Similar to the unitary operator compilation setting~\cite{Madden2022BestProblems}, we observe that the VCBE circuits saturate theoretical lower bounds on circuit parameter counts that are tied  to the degrees of freedom of the target matrix $A$. Furthermore, generic PQCs may be inefficient for matrices with inherent structure. We design specialized VCBE ansatze that are restricted to real-valued or hermitian targets and show that they reduce resource costs. For problems with known symmetries, we introduce symmetry-restricted VCBE ansatze, investigating their potential to streamline the optimization process and produce more compact quantum circuits. Additionally, we find indications that the complexity of the symmetric ansatz circuits is correlated to the algebraic structure of its circuit generators.

Our research provides a deeper understanding of the potentials and limitations of VCBE and offers practical methodologies for constructing block-encodings with significantly reduced resource overhead compared to standard techniques. Our findings highlight the utility of variational methods not just for hybrid algorithms, but as a powerful tool for realizing essential quantum algorithmic primitives efficiently.\\

This paper first reviews the framework applying Variational Quantum Compilation methods to VCBE in Section \ref{section2}. Then, in Section \ref{section3}, the generic ansatz circuits are motivated, constructed and analyzed in terms of expressivity and trainability. Section \ref{section4} goes into detail on the symmetric ansatze and their performance before Section \ref{section5} benchmarks our optimized ansatze against state-of-the art block-encoding circuits.
\section{Framework of Variational Compilation of Block-Encodings}
\label{section2}

Introduced in~\cite{Kikuchi2023RealizationComputer}, we will first review the framework of Variational Compilation of Block-Encodings in this section.
This work will deal with inputs of the form $A \in \mathbb{C}^{2^n \times 2^n}$. Arbitrary $A$ are constructed by zero-padding. In general, block-encodings are unitary encodings of non-unitary matrices. A given input matrix $A$ is encoded into a sub-block of the $N=n+m$ qubit block-encoding unitary $U_\mathrm{A}$ via
\begin{equation}
U_\mathrm{A} = \begin{bmatrix} 
            A/\alpha & \ast\\
            \ast & \ast
        \end{bmatrix},
\end{equation}
where the matrix $A$ is rescaled by the sub-normalization factor $\alpha \in \mathbb{R}$, is chosen to be encoded into the $\dyad{0^m}{0^m}$ sub-space of the $m$-qubit ancilla register. To keep our notation consistent we denote the sub-space of all ancillas being in the zero state by $\bra{0^m} \equiv \bra{0}^{\otimes m}$. Since $U_\mathrm{A}$ has to be unitary, the only requirement on $A/\alpha$ is that the spectral norm
\begin{equation}
   \norm{A/\alpha}_2 \leq 1 \, ,
   \label{eq:norm}
\end{equation}
in order to preserve the norm of an input state that $U_\mathrm{A}$ operates on.
While the specific values outside of the $\dyad{0^m}{0^m}$ subspace are constrained by the required unitarity of $U_\mathrm{A}$, they are not unique to a target $A$.\\
An efficient construction of unitary operators such as $U_\mathrm{A}$ is important in order to reduce overall resource requirements. Compilation methods~\cite{Khatri2018Quantum-assistedCompiling,Madden2022BestProblems} to implement unitary operators are well studied and are essential for running algorithms not only in the NISQ era but also beyond. While a plethora of different options exist in this field, we will focus on the numerical optimization of PQCs and adaptive circuit synthesis~\cite{Rakyta_2022,Davis2020TowardsSynthesis}.
In contrast to unitary compilation,
the target operator in VCBE is not the complete unitary $U_\mathrm{A}$, but rather just the input $A/\alpha$. The cost function is therefore chosen as
\begin{equation}
    C(\bm{\theta}) = \norm{A/\alpha - A_\mathrm{var}(\bm{\theta})}_\mathrm{F} = \epsilon \, ,
\end{equation}
calculated using the Frobenius norm $\norm{\cdot}_\mathrm{F}$, defining the encoding error $\epsilon$. $A_\mathrm{var}(\bm{\theta})$ is extracted from a variational circuit $U_\mathrm{var}(\bm{\theta})$ by projecting onto the $\dyad{0^m}{0^m}$ ancilla subspace using
\begin{equation}
    A_\mathrm{var}(\bm{\theta}) = (\bra{0^m} \otimes \mathds{1}^{\otimes n}) U_\mathrm{var}(\bm{\theta}) (\ket{0^m} \otimes \mathds{1}^{\otimes n}) \, ,
\end{equation}
where $\bm{\theta}$ are the variational parameters of the circuit $U_\mathrm{var}(\bm{\theta})$.

The circuits $U_\mathrm{var}(\bm{\theta})$ used in this work are PQCs of the form
\begin{equation}
    U_\mathrm{var}(\bm{\theta}) = \prod_l U_l(\bm{\theta}_l) \, ,
\end{equation}
with $l$ independent layers of not necessarily identical $U_{l}$ and parameters $\bm{\theta} = \{\bm{\theta}_{l}\}$. More specific $U_\mathrm{var}$ tackling different targets $A$ are constructed in the following sections. The goal of the variational compilation process is to find a set of parameters
\begin{equation}
    \bm{\theta}^\ast = \underset{\bm{\theta}}{\mathrm{argmin}} \: C(\bm{\theta})
\end{equation}
which minimize the cost function $C(\bm{\theta})$. Among different classical optimizers, we empirically found that the numerical optimization performed well with the BFGS-optimizer since optimization landscapes in VCBE were found to be smooth and issues with local minima were non-existent. Convergence is reached when the gradient norm $\nabla_{\bm{\theta}} C(\bm{\theta}) \leq 10^{-5}$, equivalent to~\cite{Kikuchi2023RealizationComputer}.
\section{Parameter-efficient generic ansatz structures}
\label{section3}

Having outlined the framework behind VCBE we now establish a baseline for the performance of its by first constructing and analyzing generic, hardware-efficient PQCs. These are very relevant in the NISQ era due to their adaptation to the gate sets and architecture of the underlying quantum hardware. While simple in construction, they have been studied extensively in variational compilation of unitary operators~\cite{Khatri2018Quantum-assistedCompiling,Madden2022BestProblems,Rakyta_2022}. It was conjectured that specific types of hardware-efficient PQCs, which we also employ in our work, span the complete space of unitary matrices from SU$(2^N)$, operating on $N$ qubits~\cite{Madden2022BestProblems}. This makes it theoretically possible to implement any block-encoding unitary $U_\mathrm{A}$. The minimal number of variational parameters in the quantum circuit and thus quantum resources needed for compilation is directly related to the degrees of freedom of a unitary matrix. To compile a general unitary a PQC needs at least $4^N-1$ independent parameters. Since VCBE does not optimize towards a target unitary, but rather a non-unitary sub-block $A$, a central question is whether VCBE is also governed by analogous resource bounds determined by the degrees of freedom of the target matrix $A$. 

High expressivity usually correlates with an exponentially hard optimization~\cite{Larocca_2025}. We aim to lower the optimization complexity by matching the ansatz circuit to properties of the target matrix in case it is hermitian and/or real-valued. Effectively this reduces the number of circuit parameters.

In the following, we first define the generic PQCs, then we discuss the theoretical bounds on parameters before finally presenting numerical results that empirically show modified lower resource bounds in VCBE.

\subsection{Variational Ansatze for Arbitrary Input Matrices}
    \label{sec:standlayan}
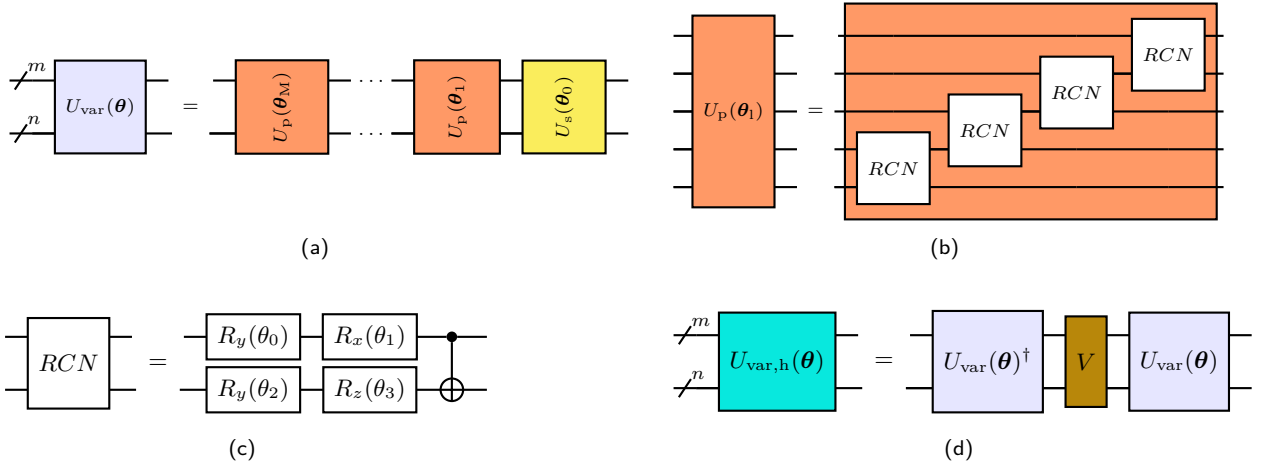
\begin{figure*}[htbp] 
    \centering
    \begin{minipage}[b]{0.48\textwidth}
        \subfloat[\label{fig:sub_uvar}]{%
            \centering 
            \raisebox{1.5cm}{%
            \begin{quantikz}[row sep = {0.7cm,between origins}, column sep = 0.3cm,font=\scriptsize]
                & \qwbundle{m} & \gate[2,style={fill=blue!10}]{U_\mathrm{var}(\bm{\theta})} & \qw \midstick[4,brackets=none]{=} & \gate[2,style={fill=atomictangerine},label style={rotate=90}]{U_\mathrm{p}(\bm{\theta}_\mathrm{M})} & \qw \ \ldots \ & \gate[2,style={fill=atomictangerine},label style={rotate=90}]{U_\mathrm{p}(\bm{\theta}_\mathrm{1})} & \gate[2,style={fill=corn},label style={rotate=90}]{U_\mathrm{s}(\bm{\theta}_\mathrm{0})} & \qw \\
                & \qwbundle{n} & \qw & \qw & \qw & \qw \ \ldots \ & \qw & \qw & \qw
            \end{quantikz}
            }
        }
    \end{minipage}
    \hfill
    \begin{minipage}[b]{0.48\textwidth}
        \subfloat[\label{fig:sub_up}]{%
            \centering
            \begin{quantikz}[row sep = {0.5cm,between origins}, column sep = 0.25cm,font=\scriptsize]
                & \gate[5,style={fill=atomictangerine}]{U_\mathrm{p}(\bm{\theta}_\mathrm{l})} & \qw \midstick[5,brackets=none]{=} & \qw \gategroup[5,steps=4,style={inner sep=1pt,fill=atomictangerine},background]{} & \qw & \qw & \gate[2]{RCN} & \qw \\
                & \qw & \qw & \qw & \qw & \gate[2]{RCN} & \qw & \qw \\
                & \qw & \qw & \qw & \gate[2]{RCN} & \qw & \qw & \qw \\
                & \qw & \qw & \gate[2]{RCN} & \qw & \qw & \qw & \qw \\
                & \qw & \qw &\qw  & \qw & \qw & \qw & \qw 
            \end{quantikz}
        }
    \end{minipage}
    \par\vspace{0.5cm}
    \begin{minipage}[b]{0.48\textwidth}
        \subfloat[\label{fig:sub_rcn}]{%
            \centering
            \begin{quantikz}[row sep = {0.7cm,between origins}, column sep = 0.3cm,font=\small]
                & \gate[2]{RCN} & \qw \midstick[2,brackets=none]{=} & \gate{R_y(\theta_0)} & \gate{R_x(\theta_1)} & \ctrl{1} & \qw\\
                & & \qw & \gate{R_y(\theta_2)} & \gate{R_z(\theta_3)} & \targ{} & \qw
            \end{quantikz}
        }
    \end{minipage}%
    \hfill
    \begin{minipage}[b]{0.48\textwidth}
        \subfloat[\label{fig:sub_uvarh}]{%
            \centering
            \begin{quantikz}[row sep = {0.7cm,between origins}, column sep = 0.3cm,font=\small]
                & \qwbundle{m} & \gate[2,style={fill=brightturquoise}]{U_\mathrm{var,h}(\bm{\theta})} & \qw \midstick[4,brackets=none]{=} & \gate[2,style={fill=blue!10}]{U_\mathrm{var}(\bm{\theta})^\dagger} & \gate[2,style={fill=darkgoldenrod}]{V} & \gate[2,style={fill=blue!10}]{U_\mathrm{var}(\bm{\theta})} & \qw \\
                & \qwbundle{n} & \qw & \qw & \qw & \qw & \qw & \qw
            \end{quantikz}
        }
    \end{minipage}
    \caption{Generic quantum circuits for Variational Compilation of Block-Encodings (VCBE). \textbf{(a):} Layered circuit of the generic ansatz $U_\mathrm{var}(\bm{\theta})$. \textbf{(b):} Example of single layer of a parameterized $U_\mathrm{p}(\bm{\theta_l})$ used in a $N=5$ qubit circuit. All RCN blocks have independent parameters. \textbf{(c):} The RCN (Rotation-CNOT) two-qubit building block. \textbf{(d):} Circuit of a Hermitian ansatz $U_\mathrm{var,h}(\bm{\theta})$. The $m$ upper qubits are ancillas and the $n$ lower qubits are system qubits.}
    \label{fig:all_circuits}
\end{figure*}

    The generic variational circuits we use have the characteristic that each layer has an identical gate structure ($U_{l} = U_\mathrm{p}$ for all $l$), but individually optimizable parameterized gates. The complete $N$-qubit circuit unitary has the form
    \begin{equation}
        U_\mathrm{var}(\bm{\theta}) = U_\mathrm{s}(\bm{\theta_0}) \prod_{l=1}^M U_\mathrm{p}(\bm{\theta}_l),
        \label{eq:standardlay}
    \end{equation}
    with a total of $M$ layers. $U_\mathrm{s}(\bm{\theta_0}) = R_0 \otimes R_1 \otimes \cdots \otimes R_\mathrm{N}$ is an appended layer consisting of single qubit rotations 
    \begin{multline}
        R_{j} = R(\vartheta_{j},\phi_{j},\lambda_{j})\\ = \begin{bmatrix}
            \cos(\vartheta_{j}/2) & -{e}^{i \lambda_{j}}\sin(\vartheta_{j}/2)\\
            \mathrm{e}^{i \phi_{j}} \sin(\vartheta_{j}/2) & \mathrm{e}^{i (\phi_{j}+\lambda_{j})} \cos(\vartheta_{j}/2)
        \end{bmatrix},
        \label{eq:singleq}
    \end{multline}
    with the three rotation angles $(\vartheta_{j},\phi_{j},\lambda_{j})$ on each of the $N$ qubits in the circuit. The circuit is depicted in Figure \ref{fig:sub_uvar}.
    
    Based on findings of parameter-efficient PQCs in~\cite{Madden2022BestProblems}, we design a circuit layer $U_\mathrm{p}(\bm{\theta}_\mathrm{l})$ as shown in Figure \ref{fig:sub_up}. It is composed from independently parameterized two-qubit $RCN$ circuits given in Figure \ref{fig:sub_rcn}. When assembled in a larger circuit this building block has the optimal ratio of free parameters to CNOT gates of 4:1. The single qubit unitary $U_\mathrm{s}(\bm{\theta_0})$ at the end of the ansatz ensures more free variational parameters in the circuit without the need for more two-qubit gates. Note that the circuit applies gates uniformly on all $N$ qubits without differentiating between ancilla and system qubits.

    \subsubsection{Hermitian Ansatz}
    The first restriction on the space of encodable matrices is by building a hermitian VCBE ansatz circuit. If the circuit unitary $U_\mathrm{A}$ is hermitian, then the block-encoded matrix $A$ is also hermitian, since
    \begin{equation}
        \begin{bmatrix} 
                A & \ast\\
                \ast & \ast
            \end{bmatrix} = U_\mathrm{A} = U^\dagger_\mathrm{A} = 
            \begin{bmatrix} 
                A^\dagger & \ast\\
                \ast & \ast
            \end{bmatrix}.
    \end{equation}
    Hermiticity can be achieved through the sequence
    \begin{equation}
        U_\mathrm{var,h}(\bm{\theta}) = U_\mathrm{var}(\bm{\theta}) V U_\mathrm{var}(\bm{\theta})^\dagger
        \label{eq:herm}
    \end{equation}
    where $U_\mathrm{var}(\bm{\theta})$ is the layered ansatz defined in eq.~\eqref{eq:standardlay} and $V$ is a hermitian operator that is not necessarily parameterized~\cite{Kikuchi2023RealizationComputer}. In general, $V$ must not commute with $U_\mathrm{var}(\bm{\theta})$. We choose $V = H^{\otimes (n+m)}$, mitigating cancellation of gates between $U_\mathrm{var}$ and $U^\dagger_\mathrm{var}$, for all hermitian circuits in this section. The circuit is shown in Figure \ref{fig:sub_uvarh}.

    \subsubsection{SU(\texorpdfstring{$2^N$}{2 exp N}) vs. SO(\texorpdfstring{$2^N$}{2 exp N})}
    Apart from being hermitian, block-encoded matrices are often purely real-valued. Examples can be coefficient matrices from a system of linear equations or various types of Hamiltonians. Ansatz blocks that incorporate general single qubit rotations span a needlessly large matrix space, possibly hindering optimizability. We exchange general single qubit rotations $R(\vartheta,\phi,\lambda)$ with real-valued $R_\mathrm{y}(\vartheta) = R(\vartheta,0,0)$ in $U_\mathrm{s}(\bm{\theta}_{0})$ and use only $R_\mathrm{y}$-rotations in each $U_\mathrm{p}(\bm{\theta}_{l})$. This restricts the circuit unitary to SO$(2^N)$ and aims to make encoding of target matrices that are real-valued more resource-efficient.

\subsection{Target matrices and theoretical lower parameter number bounds}
    \label{sec:Inputs}
    Before assessing the expressibility of different circuits, we analyze resource bounds. A general unitary matrix $U$ from SU($2^n$) has in total $N_\mathrm{p} = 4^n-1$ independent matrix entries and to compile this matrix exactly, a circuit using $n$ qubits has to have minimally $N_\mathrm{p}$ adjustable parameters. Unitary gate decomposition using variational methods has approached this lower bound~\cite{Rakyta_2022}.
    Our work will investigate if lower bounds are also met in the VCBE setting with non-unitary target matrices. The number of independent parameters for different general inputs is listed in table \ref{tab:N_freeP}.
    \begin{table}[h]
        \centering
        \begin{tabular}{c|c|c}
                & complex & real \\\hline
            arbitrary & $2 \cdot 4^n$ & $4^n$\\
            hermitian & $4^n$ & $\frac{2^n(2^n+1)}{2}$\\
            unitary & $4^n-1$ & $\frac{2^n(2^n-1)}{2}-1$
        \end{tabular}
        \caption{Number of free parameters $N_\mathrm{p}$ in a square $2^n \times 2^n$ input matrix.}
        \label{tab:N_freeP}
    \end{table}

    Apart from dense and unstructured matrices, sparse and/or structured matrices play an important role in quantum algorithms. As an example for a sparse and structured matrix the transverse field Heisenberg Hamiltonian
    \begin{multline}
        H_\mathrm{Heis} = \sum_{i=1}^{n-1} \left ( J_\mathrm{x} X_{i}X_{i+1} + J_\mathrm{y} Y_{i}Y_{i+1} + J_\mathrm{z} Z_{i}Z_{i+1} \right )\\ + \sum_{i=1}^{n} h X_{i},
    \label{eq:Heisenberg_H}
    \end{multline}
    with the Pauli operators $X,Y,Z$ and coefficients $J_\mathrm{x},J_\mathrm{y},J_\mathrm{z},h$. Non-periodic boundary conditions are used in this work, unless stated otherwise.

    While non-square and general $k \times l$ input matrices are not directly considered in this work, they can always be expanded to a $2^n \times 2^n$ matrix by adding zeros. All target matrices are scaled by $\alpha = \norm{A}_2 + \delta$ so that the condition in eq. \eqref{eq:norm} is met and additionally a small $\delta = 10^{-2}$ is used to avoid numerical instabilities.

\subsection{Numerical Results}
\label{sec:numres}
   This section will present the results of numerically applying the VCBE workflow to the described ansatze. It will also make connections between the degree of freedom in circuits and inputs, empirically showing that lower bounds in circuit parameters can be transferred to VCBE, resulting in efficient encodings of dense input matrices. We focus exclusively on ansatze with a single ancilla qubit ($m=1$), due to an observed drop in resource efficiency when adding additional ancillas, see Appendix \ref{app:ancN}.
   
\begin{figure*}[t]
        \centering
        \subfloat[]{\includegraphics{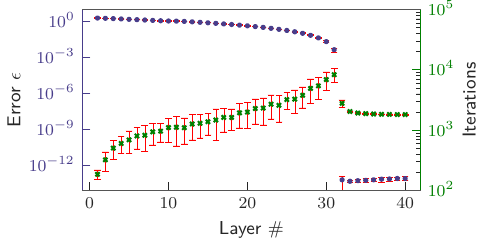}\label{fig:opt_curves}}%
        \hfill 
        \subfloat[]{\includegraphics{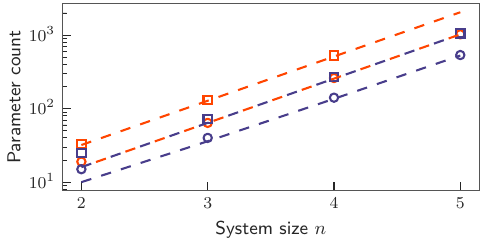}\label{fig:Pcount}}
        \caption{(\textbf{a}): Optimization performance for a $n+m = 4+1$ qubit ansatz circuit. Blue dots are the average error $\epsilon = \norm{A-A_\mathrm{var}}$ of block-encoding 10 different random matrices from a uniform distribution and initializing each with 10 different parameter sets. Green crosses indicate the number of average iterations the BFGS-optimizer needed to converge. In both cases error bars are the standard deviation. (\textbf{b}): Matrix size dependence of the parameter count $N_\mathrm{p,c}$ of optimized circuits for complex(squares), real(circles), arbitrary(orange) and hermitian(blue) input matrices with circuits at the system size respective threshold layer. Dashed lines are optimal values $N_\mathrm{p}$ from Table \ref{tab:N_freeP}. No errorbars are added since 10 probed random input matrices and 10 random initial parameters at each input optimized exactly at the same layer number.}
        \label{fig:combined}
    \end{figure*}

   Using the Jax python library~\cite{jax2018github} to calculate the circuit unitaries and JAXopt~\cite{blondel2022efficientmodularimplicitdifferentiation} running the BFGS optimizer, a general behavior of the ansatz for arbitrary target matrices is seen in Figure \ref{fig:opt_curves} with a $N = 5$ qubit ansatz encoding on $n=4$ system qubits. First of all, it shows that the optimization of a generic circuit is possible with an error reaching $\epsilon \lesssim 10^{-12}$. This will be designated to be an exact encoding since this error is limited by numerical accuracy. For multiple random input matrices and starting points in the variational landscape, the achieved encoding error first stays above $\epsilon = 10^{-3}$ and drops sharply after reaching a threshold layer at $M_\mathrm{thres} = 32$.
    Performing the optimization process for inputs that are described in Section \ref{sec:Inputs} (excluding unitary), gives the resource counts for circuits with length $M_\mathrm{thres}$ listed in Table \ref{tab:A2_res}.
    \begin{table*}[th]
        \centering
        \begin{tabular}{c|c|c|c|c}
            & \multicolumn{2}{c|}{Random Input} & \multicolumn{2}{c}{Heisenberg (sparse)}\\\hline
            Ansatz & SU($2^5$) & SO($2^5$) & SU($2^5$) & SO($2^5$)\\\hline
            & \multicolumn{4}{c}{Parameter count $N_\mathrm{p,c}$}\\\hline
            arbitrary & 527 (512) & 261 (256) & 527 & 261\\
            hermitian & 271 (256) & 141 (136) & 271 & 141\\\hline
            & \multicolumn{4}{c}{CNOT count}\\\hline
            arbitrary & 128 (125) &  128 (126) & 128 & 128\\
            hermitian & 128 (61) & 136 (66) & 128 & 136
        \end{tabular}
        \caption{Resources used to exactly compile a $N=5$ qubit block-encoding $U_\mathrm{A}$ for a $2^4 \times 2^4$ target matrix $A$. Random inputs are either complex- or real-valued and (non-)hermitian, depending on the chosen ansatz. The Heisenberg matrix is always real and hermitian, while the ansatz circuits vary. Theoretical lower bounds on parameters $N_\mathrm{p}$ (from Table \ref{tab:N_freeP}) and CNOT gates $\mathrm{TLB}_\mathrm{CNOT}$ (from Table \ref{tab:N_NlG}) are in brackets.}
        \label{tab:A2_res}
    \end{table*}
    It is observed that the number of free parameters in the quantum circuit is close to the lower bound of independent parameters that make up the target matrix. Figure \ref{fig:Pcount} demonstrates this behavior for different input sizes.\\
    We empirically show for small system sizes up to $n=5$ and $m=1$, that the lower bounds found for unitary compilation~\cite{Madden2022BestProblems}  can be extended to a setting where the target operator is block-encoding a specific input matrix. The lower bounds will only be obeyed if the ansatz circuit has the same property, meaning a non-hermitian circuit will still encode a hermitian input with the number of parameters corresponding to a non-hermitian input. The same case holds for circuits from SU($2^N$) encoding a purely real input matrix.
    Deviation from the lower bounds can be mainly attributed to the fact that variational parameters do not get added continuously, but rather in blocks for every additional layer. This causes an overhead.
    
    Together with lower bounds on variational parameters, we establish bounds on CNOT gates. The basic $RCN$-blocks combined with the general single qubit gates $U_\mathrm{s}$ construct a circuit that is optimal in terms of independent parameters~\cite{Shende2004MinimalCircuits}. 
    The total number of parameters in our ansatz for arbitrary target matrices is
    \begin{equation}
        N_\mathrm{p,c} = 4 \cdot N_{RCN} + 3 \cdot (n+1) \, ,
    \end{equation}
    where the first term is dependent on the number of $RCN$ blocks $N_{RCN}$ and the second term is the number of parameters in $U_\mathrm{s}$. Since a general $2^n \times 2^n$ complex-valued target matrix $A$ has $2 \cdot 4^n$ degrees of freedom, we can obtain for $N_\mathrm{p,c} = 2 \cdot 4^n$ and one ancilla qubit the theoretical lower bound ($\mathrm{TLB}_\mathrm{CNOT}$) on the number of CNOT gates
    \begin{equation}
        \mathrm{TLB}_\mathrm{CNOT} = \left \lceil \frac{1}{4} (2 \cdot 4^n-3(n+1)) \right \rceil \, .
        \label{eq:TLB}
    \end{equation}
    
    Generalizing to multi-qubit entangling gates, the ratio    
    \begin{equation}
        a = \frac{\mathrm{\# \: circuit \: parameters}}{\mathrm{\# \: multi\mbox{-}qubit \: gates}}
        \label{eq:aratio}
    \end{equation}
    describes the number of variational parameters a multi-qubit gate can absorb without the need to merge parameterized gates in the full circuit. 
    The ratio is $a = 4$ for the $RCN$ ansatze with CNOT gates and $a=6$ for circuits constituted of CCNOT gates and rotations analogous to the $RCN$ ansatz. Table \ref{tab:N_NlG} lists lower bounds of non-local gate counts for different input matrices. 
    \begin{table}[ht]
        \centering
        \begin{tabular}{c|c|c}
                & complex & real \\\hline
            &&\\[-6pt]
            arbitrary & $\frac{1}{a} (2 \cdot 4^n-3N)$ & $\frac{1}{a} (4^n-N)$\\[5pt]
            hermitian & $\frac{1}{a} (4^n-3N)$ & $\frac{1}{a} (\frac{2^n(2^n+1)}{2}-N)$\\[5pt]
        \end{tabular}
        \caption{Lower bounds for the number of multi-qubit-gates for different $2^n \times 2^n$ input matrices. $N$ is the total qubit number of the ansatz circuit.}
        \label{tab:N_NlG}
    \end{table}
    While these bounds are approached for the arbitrary ansatze (see Table \ref{tab:A2_res}), hermitian ansatze fail to do so. The reason is that the specific circuit structure for hermitian circuits halve the parameter per CNOT gate ratio through its particular construction. Therefore, hermitian and real ansatze do not necessarily reduce resource costs regarding the number of CNOT gates but rather simplify optimization through a smaller dimension of the parameter space.
    
\begin{figure}[h]
        \centering
        \includegraphics{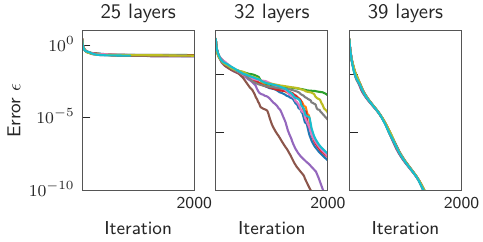}
        \caption{Encoding error during optimization runs for different layer numbers. Each color is a run starting at a 10 random initial parameter sets.}
        \label{fig:opt_lay}
    \end{figure}

    A further observation from Figure \ref{fig:opt_curves} is that the number of iterations in finding a minimum increases exponentially with the number of circuit layers, in line with the theory of barren plateaus in variational algorithms using generic ansatz circuits~\cite{Larocca_2025}. When reaching the threshold for theoretical exactness defined by the $\mathrm{TLB}_\mathrm{CNOT}$ in eq. \eqref{eq:TLB}, the convergence behavior changes markedly. When the ansatz is overparameterized, the number of optimization steps is significantly reduced and also independent of circuit depth. This effect is also seen in Figure \ref{fig:opt_lay} where barren plateaus still appear for $M = M_\mathrm{thres}$ but vanish for larger parameter spaces. An analysis of the optimization landscape for unitary compiling~\cite{Madden2022BestProblems} has similar results including the conjecture that a global minimum can be reached from any point on the optimization landscape when in the overparameterized regime. While a more sophisticated analysis is not in the scope of this work, our findings suggest that the conjecture is still upheld in the case of VCBE.
    When implementing these findings in a practical VCBE scheme, it is preferable to optimize toward an optimal layer $M_\mathrm{thres}$ from above and stop reducing the number of layers only when exact encoding becomes unsuccessful. The threshold layer is calculated by 
    \begin{equation}
        M_\mathrm{thres} = \left \lceil \frac{N_\mathrm{p} - 3N}{a \cdot N_\mathrm{nlg,c}} \right \rceil \, ,
    \end{equation}
    where $N_\mathrm{nlg,c}$ is the number of non-local gates in the circuit.

    The findings from the exemplary ansatz used over the course of this section are supported by Appendix \ref{app:AddGenCircs}, where we evaluated further ansatz circuits. Their ability to encode a target matrix is almost identical. We can identify necessary conditions for an exact encoding of a random complex/real, (non-)hermitian input matrix, which are condensed to:
    \begin{itemize}
        \item Input $A$ has $\norm{A/\alpha}_2 \leq 1$.
        \item All qubits can be entangled with each other. Failure on this part will result in an insufficient ansatz expressibility.
        \item The number of independent parameters in the ansatz must be larger or equal to independent parameters of the input. 
        \item The ansatz has to be in the correct operator space (for example, a $SO(2^n)$ ansatz cannot encode an arbitrary complex matrix).
    \end{itemize}
    If all conditions are met, variationally encoding an input exactly was found to be surprisingly robust in the over-parameterized regime.
    
    It is important to note that encoding a sparse or structured input generally does not reduce circuit resources. When encoding the matrix of a Heisenberg Hamiltonian, even though the spin sites and ansatz gates are ordered in a linear fashion, no difference can be observed in resource requirements (see Table \ref{tab:A2_res}). For special cases like diagonal matrices, a difference could be observed, but adjusting generic ansatze to fit a specific sparsity pattern is highly non-trivial and will be considered in future work.

\section{Parameter reduction with symmetric ansatz structures for symmetric inputs}
\label{section4}

Utilizing that the target matrix is either real and/or hermitian was the first step in showing that additional information about the input and its incorporation into the ansatz circuit can effectively reduce the ansatz complexity and enhance trainability. However, most input data is more structured and exhibits symmetries, particularly in physical Hamiltonians. Generic ansatz constructions have a high expressivity leading to efficient encodings of dense and unstructured inputs, but also significant overheads when dealing with inputs that are more structured. This was clearly observed when attempting to encode a Heisenberg Hamiltonian in the previous section. A proven method to tackle an excess of expressivity in Variational Quantum Eigensolvers~\cite{Gard_2020,Lacroix_2023} and Quantum Machine Learning circuits~\cite{Meyer_2023,Larocca_2022} is to design symmetric ansatz circuits. In this section we develop an approach to translate input symmetries to ansatz symmetries and explore different examples in terms of circuit structures and trainability. Additionally, we make a case for determining the ansatz expressibility from the set of circuit generators.

\subsection{From Input Symmetry to Ansatz Symmetry}
    In the following, the input matrix is a Hamiltonian $H$ which is assumed to have a spectral norm $\norm{H/\alpha}_2 \leq 1$, which can always be established by choosing a suitable sub-normalization $\alpha$. It is important to point out that the presented methods are also applicable to more general inputs, including non-hermitian matrices. First, we will give an intuitive approach for creating ansatz circuits, formalizing it as we go.
    
    Given a symmetry operation $S$, the Hamiltonian is invariant under $S$ if
    \begin{equation}
        [H,S] = 0.
    \end{equation}
    A $n+1$ qubit block-encoding $U_\mathrm{H}$ of $H$ should therefore have the restriction    
    \begin{equation}
        [(\bra{0} \otimes \mathds{1}^{\otimes n})U_\mathrm{H}(\ket{0} \otimes \mathds{1}^{\otimes n}),S] = 0
        \label{eq:geninvariant}
    \end{equation}
    or
    \begin{equation}
        \label{eq:U_H_S_Tilde}
        [U_\mathrm{H},\Tilde{S}] = \begin{bmatrix}
        [H,S] & \ast\\
        \ast & \ast
        \end{bmatrix} = \begin{bmatrix}
        0 & \ast\\
        \ast & \ast
        \end{bmatrix} \, ,
    \end{equation}
    with $\Tilde{S}$ being a $n+1$ qubit operator containing the symmetry $S$.
    Since we want to find an explicit circuit for $U_\mathrm{H}$, let us now find a suitable $\Tilde{S}$.\\
    Using 
    \begin{equation}
        \Tilde{S} = (\mathds{1} \otimes S) = \begin{bmatrix}
            S & 0\\
            0 & S
        \end{bmatrix}
    \end{equation}
    to apply the symmetry operation only on the system register, we get that
    \begin{equation}
        [U_\mathrm{H},\Tilde{S}] = \begin{bmatrix}
        [H,S] & [\ast,S]\\
        [\ast,S] & [\ast,S] \, ,
        \end{bmatrix}
    \end{equation}
    showing that equation \eqref{eq:U_H_S_Tilde} is fulfilled by our choice of $\Tilde{S}$. It is straightforward to generalize this specific $\Tilde{S}$ to $n+m$ qubit block-encodings, since $\Tilde{S} = \mathds{1}^{\otimes m} \otimes S$ acts only on the system register non-trivially.

\subsection{Symmetric Circuit Construction}
   We now construct a quantum circuit invariant under $\Tilde{S}$. To that end, this section first discusses a method for constructing a set of symmetric circuit generators and secondly a heuristic for selecting from this set and deciding if a target matrix is implementable.

    \subsubsection{Circuit Layout}
    \label{sec:SymmCircs}
\begin{figure*}
        \centering
        \begin{quantikz}[row sep = 0.4cm, column sep = 0.5cm,font=\scriptsize]
            & \qw & \gate[2,style={fill=blue!10}]{U_\mathrm{var}(\bm{\theta})} & \qw \midstick[2,brackets=none]{=} & \qw & \gate[style={fill=caribbeangreen}]{R(\vartheta_0,\phi_0,\lambda_0)} & \ctrl{1} \gategroup[2,steps=2,style={dashed,rounded corners, inner xsep=2pt},background, label style={label position=below right,anchor=north,yshift=-0.2cm}]{{$\times M$}} & \gate[style={fill=caribbeangreen}]{R(\vartheta_\mathrm{i},\phi_\mathrm{i},0)} & \qw\\
            & \qwbundle{n} & \qw & \qw & \qwbundle{n} & \qw & \gate[style={fill=capri}]{P_{i}(\bm{\theta}_{i})} & \qw  & \qw
        \end{quantikz}
        \caption{GQSP-type ansatz. The upper qubit is the ancilla qubit and the $n$ lower qubits are the system qubits.}
        \label{fig:GQSPAnsatz}
    \end{figure*}
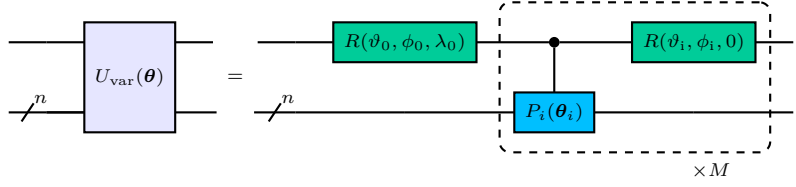

    Since $\Tilde{S}$ is a tensor product, a $n$ qubit unitary operator $P$ applied to the system register will fulfill $[\mathds{1} \otimes P,\Tilde{S}] = 0$ if $[P,S] = 0$. However, sequences of unitary operators on the system register separated from the ancillas are also unitary and thus will not be able to block-encode a $2^n \times 2^n$ non-unitary matrix. A possible solution lies in turning $P$ into a controlled version    
    \begin{equation}
        cP = \dyad{0}{0} \otimes \mathds{1} + \dyad{1}{1} \otimes P = \begin{bmatrix}
            \mathds{1}^{\otimes n} & 0\\
            0 & P
        \end{bmatrix}
    \end{equation}
    which is still invariant under $\Tilde{S}$ with
    \begin{equation}
        [cP,\mathds{1} \otimes S] = \begin{bmatrix}
            [\mathds{1}^{\otimes n},S] & 0\\
            0 & [P,S]
        \end{bmatrix} = 0.
        \label{eq:cP1S}
    \end{equation}
    Inspired by the Generalized QSP algorithm~\cite{Motlagh2024}, a sequence of not necessarily identical $cP_{i}$ operators, or their parameterized version $cP_{i}(\bm{\theta}_{i})$ can be interspersed with universal single qubit rotations on the ancilla qubit. The single qubit rotation will not affect the symmetry on the system qubits and are defined by eq. \eqref{eq:singleq}. The symmetric ansatz takes the form
    \begin{equation}
        U_\mathrm{var}(\bm{\theta}) = \left [ \prod_{i=1}^M R(\vartheta_{i},\phi_{i},0) cP_{i}(\bm{\theta}_{i}) \right ] R(\vartheta_{0},\phi_{0},\lambda_{0}) \, ,
        \label{eq:sym_ansatz_seq}
    \end{equation}
    with parameters $\bm{\theta} = \{\vartheta_{i},\phi_{i},\lambda_{i},\bm{\theta}_{i}: \: \forall i=0,1,...,M\}$ and is presented as a circuit in Figure \ref{fig:GQSPAnsatz}. All $R_\mathrm{z}$ rotations conditioned by $\lambda$ can be moved through the controls so that $\lambda_{i}=0 \: \forall i \in \{ 1,...,M \}$. A hermitian form of this ansatz is constructed easily by taking the complex conjugate of $U_\mathrm{var}(\bm{\theta})$ just as in eq. \eqref{eq:herm} and inserting a single Hadamard gate $V = H \otimes \mathds{1}^{\otimes n}$ on the ancilla qubit.

    \subsubsection{Generating Sets}
    The next step is to find symmetry restricted operators $P_{i}$ that are able to encode the Hamiltonian. 
    We will denote a Pauli string as
    \begin{equation}
        \chi = \bigotimes^n_{j=0} \sigma^{(j)} = \sigma^{(0)} \otimes \sigma^{(1)} \otimes \cdots \otimes \sigma^{(n-1)}
    \end{equation}
    with the Pauli operators $\sigma^{(j)} \in \{\mathds{1},X,Y,Z\}$. If the context is clear, the tensor products are omitted. All possible permutations of Pauli strings of length $n$ build up a basis of the Lie algebra $\mathfrak{su}(2^n)$ associated to the special unitary group $\mathrm{SU}(2^n)$. The Lie algebra is defined as
    \begin{equation}
        \mathfrak{su}(2^n) = \mathrm{span}_\mathbb{R} \left \{ i \bigotimes_{j=0}^n \sigma^{(j)} \right \} \, \backslash \, \left \{ i\mathds{1}^{\otimes n} \right \} \, .
        \label{eq:su_span}
    \end{equation}    
    Adopting the notation from~\cite{Mansky_2025} and generalizing the single symmetry operation $S$ to the case of a subset $\mathcal{M} \subseteq M_{2^n}(\mathbb{C})$ that contains any desired symmetry, an invariant Lie algebra is defined as 
    \begin{equation}
        \mathcal{M}\mathfrak{su}(2^n) = \{ a \in \mathfrak{su}(2^n) : SaS^{-1} = a, \: \forall S \in \mathcal{M} \} \, .
    \end{equation}
    If $\mathcal{M}\mathfrak{su}(2^n)$ is a connected Lie algebra, the exponential map
    \begin{equation}
        \mathcal{M}\mathfrak{su}(2^n) \rightarrow \mathcal{M}\mathrm{SU}(2^n)
    \end{equation}
    is surjective and thus any element of the corresponding Lie group $\mathcal{M}\mathrm{SU}(2^n)$ can be generated by the symmetric Lie algebra. In turn, the Lie group is
    \begin{equation}
        \mathcal{M}\mathrm{SU}(2^n) = \{ A \in \mathrm{SU}(2^n) : AS = SA, \: \forall S \in \mathcal{M} \} \, ,
    \end{equation}
    which means its elements are also invariant under the symmetries from $\mathcal{M}$. Therefore, by choosing a generator $G \in \mathcal{M}\mathfrak{su}(2^n)$ we can ensure that $P \in \mathcal{M}\mathrm{SU}(2^n)$ with
    \begin{equation}
        P = \mathrm{e}^{\theta G}.
    \end{equation}
    The dimensionality of $\mathcal{M}\mathfrak{su}(2^n)$ can be significantly lower compared to the full Lie algebra and depends on the symmetries in $\mathcal{M}$.
    
\subsection{Heuristics for Generator Selection and Optimization}

    \subsubsection{Generator Selection}
    \label{sec:GenSel}
    Finding the symmetric Lie algebra $\mathcal{M}\mathfrak{su}(2^n)$ is done on a case-by-case basis. Once achieved, a set of circuit generators $\mathcal{G} = \{G_{i}\}$ with $G_{i} \in \mathcal{M}\mathfrak{su}(2^n)$ has to be chosen. In our work, this is largely done through a heuristic selection, adjusting $\mathcal{G}$ depending on the success of an exact optimization of the target input. Within the heuristic, two strategies are useful in finding a set $\mathcal{G}$ that is able to encode any given Hamiltonian efficiently:
    \begin{itemize}
        \item \textbf{Hardware based implementability}\\
            The circuits realizing each $P_{i}(\bm{\theta}_{i})$ have to be exact implementations of the operator in order to be fully symmetry restricted and in the best case easily implementable. Take for example a 3+1 qubit ansatz that should obey a full permutation symmetry. Two possible generators from the symmetric Lie algebra are
            \begin{equation*}
                G_1 = i(ZZI+ZIZ+IZZ)
            \end{equation*}
            and
            {\small
            \begin{equation*}
                G_2 = i(ZYI+YZI+ZIY+YIZ+IZY+IYZ) \, .
            \end{equation*}}
            While $G_1$ can be easily decomposed and implemented as
            \begin{equation*}
                P = \mathrm{e}^{i \theta (ZZI+ZIZ+IZZ)} = \mathrm{e}^{i \theta ZZI} \mathrm{e}^{i \theta ZIZ} \mathrm{e}^{i \theta IZZ} \, ,
            \end{equation*}
            $G_2$ can only be implemented by approximation, for example through Trotterization (in case $G_2$ is not native to the hardware) since the composing Pauli strings do not commute. Excluding generators requiring complex decompositions can of course lower the expressivity of the circuit, but we will empirically show in Section \ref{sec:Symm_res} that the full basis of $\mathcal{M}\mathfrak{su}(2^n)$ is often not necessary to encode a symmetric target matrix.
        \item \textbf{Matrix expressibility}
            So far we have defined how a symmetric circuit should be composed to implement a symmetric matrix. Alternatively, we can turn this formulation around and ask what $H$ can be constructed by choosing a specific set of generators $\mathcal{G}$. This means one has to characterize the composition of the variational matrix block
            \begin{equation}
                F = (\bra{0} \otimes \mathds{1}^{\otimes n}) U_\mathrm{var} (\ket{0} \otimes \mathds{1}^{\otimes n})
            \end{equation}
            which evaluates to
            \begin{equation}
                F = \prod_{i=1}^{M} \left ( a_{i} \mathds{1}^{\otimes n} + b_{i} P_{i}(\theta_{i}) \right ) \mathrm{e}^{i\lambda_0} \, .
                \label{eq:Fprod}
            \end{equation}
            The coefficients $a_{i}, \: b_{i} \in \mathbb{C}$ and $\lambda_0 \in \mathbb{R}$ are determined by the single qubit rotations on the ancilla. On the other hand, we can expand $F$ in a basis $\mathcal{B}$ as
            \begin{equation}
                F = \sum_{k} f_{k} B_{k},
            \end{equation}
            with complex coefficients $f_{k} \in \mathbb{C}$ and linearly independent operators $B_{k} \in \mathcal{B}$. We can construct a basis $\mathcal{B}_\mathrm{G}$ from the set $\mathcal{G}$ by first calculating the basis elements of the dynamical Lie algebra of $\mathcal{G}$ and subsequently the associative closure of that set, details are provided in Appendix \ref{app:Bcalc}. We conjecture that
            \begin{equation}
                F \in \mathrm{span}_\mathbb{C}(\mathcal{B}_\mathrm{G}) \, ,
                \label{eq:Fspan}
            \end{equation}
            making it possible to check the implementability of a given $H$ without optimizing the ansatz.
    \end{itemize}

    \subsubsection{Ansatz Optimization}
    Circuit expressibility and trainability depend on the structure of the ansatz circuit. As a heuristic, choosing the sequence of operator blocks $cP_{i}$ can be done in the following ways:
    \begin{itemize}
        \item \textbf{Random layering}\\
            In its simplest form, the ansatz is built up layer by layer while choosing the generators $G_i \in \mathcal{G}$ at random until a sufficiently small encoding error is satisfied. Alternatively, the threshold layer for exact encoding can be reached by generating long ansatz circuits and removing layers until achieving the desired accuracy/circuit depth. When a good estimate for the threshold layer can be made, the latter strategy is more efficient analogous to the discussion in Section \ref{sec:numres}. A meaningful estimate for the threshold layer number is calculated with the method from Appendix \ref{app:Bcalc}.
        \item \textbf{Search algorithms}\\
            The selection process can be formulated in terms of a tree search problem where after each new layer, all possible generators are tested and depending on the optimization error, specific generators are used. Unitary compilation has had high success rates using these methods in finding low gate-count solutions for complex operators~\cite{younis2020qfastquantumsynthesisusing,Davis2020TowardsSynthesis}. The search algorithms perform a more systematic approach to compilation but in turn have much longer run-times since multiple circuit structures have to be evaluated at every step.
        \item \textbf{Random/sequential parameter initialization}
            Two different possibilities arise when setting the initial parameters of the ansatz circuit. On the one hand a random initialization is straightforward and promises good results for convex optimization problems. Alternatively, a faster optimization is possible by re-using the previously optimized parameters and initializing the new layer as unity with $\bm{\theta} = \bm{0}$. If the distance to the target operator is already small, a further step will only slightly change the circuit parameters.
    \end{itemize}

\subsection{Results}
    \label{sec:Symm_res}

    \subsubsection{Input Matrix and Symmetries}
    \label{sec:inputs_syms}
     We focus on a system described by the Heisenberg Hamiltonian defined by eq. \eqref{eq:Heisenberg_H}. The Hamiltonian is invariant under the symmetry operator $S = X^{\otimes n}$ making it $Z_2$ symmetric. Additionally, the spins can be geometrically arranged in different ways and thus allow for various symmetries. We will deal with three of them:    
    \begin{itemize}
        \item Reflection symmetry, $Z_\mathrm{2,xz}$\\
         In case of non-periodic boundary conditions, the spin sites are arranged on a straight chain and the Hamiltonian is invariant under reflection in the middle of the chain. It is denoted as $Z_\mathrm{2,xz}$ since it is a reflection in the $xz$-plane compared to the general $Z_2$-symmetry in the $zy$-plane.
        \item Cyclic symmetry, $C_\mathrm{n}$\\
        In the case of periodic boundary conditions a circular chain with $n$ sites has a $C_\mathrm{n}$ symmetry.
        \item Permutation symmetry, $S_\mathrm{n}$\\
        The system is invariant under the exchange of spin sites.
    \end{itemize}

    We restrict ourselves to always making use of the $Z_2$ symmetry and one of either $Z_\mathrm{2,xz}$, $C_\mathrm{n}$ or $S_\mathrm{n}$ symmetry.
    Apart from these geometric symmetries, the input matrix is hermitian. Ansatz circuits are therefore hermitian as described in Section \ref{sec:SymmCircs}.

    \subsubsection{Generator Sets}
    \label{sec:GenSets}
    Choosing suitable circuit generators is crucial for the ability to encode an input matrix. We use the heuristics for generator selection from Section \ref{sec:GenSel}. While they allow for a range of different options, we choose only operators corresponding to terms making up the Hamiltonian. For the different symmetries, the generating sets $\mathcal{G}$ of an exemplary 4-qubit system are:
        
    {\footnotesize
    \begin{align*}
        \mathcal{G}_\mathrm{Z_{2,xz}} = \{ & i(XXII + IIXX), iIXXI,\\
                                           & i(YYII + IIYY), iIYYI,\\
                                           & i(ZZII + IIZZ), iIZZI,\\
                                           & i(XIII + IIIX), i(IXII + IIXI)\},
    \end{align*}
    }
    {\footnotesize
    \begin{align*}
        \mathcal{G}_\mathrm{C_4} = \{ & i(XXII + IXXI + IIXX + XIIX),\\
                                      & i(YYII + IYYI + IIYY + YIIY),\\
                                      & i(ZZII + IZZI + IIZZ + ZIIZ),\\
                                      & i(XIII + IXII + IIXI + IIIX)\},
    \end{align*}
    }
    {\footnotesize
    \begin{align*}
        &\mathcal{G}_\mathrm{S_4} = \\
                                    \{ & i(XXII + IXXI + IIXX + XIIX + IXIX + XIXI),\\
                                       & i(YYII + IYYI + IIYY + YIIY + IYIY + YIYI),\\
                                       & i(ZZII + IZZI + IIZZ + ZIIZ + IZIZ + ZIZI),\\
                                       & i(XIII + IXII + IIXI + IIIX)\}
    \end{align*}
    }
    It is straightforward to construct generator sets for other qubit numbers.
    All resulting $P_{i}$ are easy to implement because the Pauli strings making up a generator commute with each other. The controlled version, $cP_{i}$, is built by using Pauli gadgets as shown in Appendix \ref{app:GQSPcircuits}.

    \subsubsection{Numerical Results}
    The symmetric ansatz circuits demonstrate input encoding that achieves a significantly better parameter efficiency compared to generic circuits. As shown in Table \ref{tab:GQSPres}, the symmetric circuit constructions reliably encode Heisenberg Hamiltonians with random coefficients once a critical threshold of layers is exceeded, enabling exact representation. The number of required parameters is reduced compared to the full $N_\mathrm{p} = 2 \cdot 4^n$ for general matrices ($N_\mathrm{p} = 4^n$ in the Hermitian case). The results were obtained through using the \textit{random layering} protocol and \textit{symmetry-adapted search algorithm}.
    \begin{table*}[t]
        \centering
        \begin{tabular}{c|ccc|ccc|ccc}
            & \multicolumn{3}{c|}{$Z_\mathrm{2,xz}$} & \multicolumn{3}{c|}{$C_\mathrm{n}$} & \multicolumn{3}{c}{$S_\mathrm{n}$}\\\hline
            n & $D$ & $M_\mathrm{thres}$ & $N_\mathrm{p,c}$ & $D$ & $M_\mathrm{thres}$ & $N_\mathrm{p,c}$ & $D$ & $M_\mathrm{thres}$ & $N_\mathrm{p,c}$\\\hline
            2 & 6 & 3\textbar2 & 12\textbar9 \:   &   6 & 3\textbar2 & 12\textbar9 \: & 6 & 3\textbar2 & 12\textbar9 \:\\
            3 & 20 & 7\textbar7 & 24\textbar24 &   10 & 4\textbar4 & 15\textbar15 & 10 & 4\textbar4 & 15\textbar15\\
            4 & 72 & 25\textbar24 & 78\textbar75 & 28 & 9\textbar9 & 30\textbar30 & 19 & 6\textbar6 & 21\textbar21\\
            5 &    &       &       & 68 & \: -\textbar23 & \: -\textbar72 & 28 & -\textbar9 & \: -\textbar30\\
            6 &    &       &       &     &     &      & 44 & \: -\textbar15 & \: -\textbar48\\
            7 &    &       &       &     &     &      & 60 & \: -\textbar20 & \: -\textbar63\\
            8 &    &       &       &     &     &      & 85 & \: -\textbar28 & \: -\textbar87
        \end{tabular}
        \caption{Characteristic values for the optimization of symmetric ansatze exactly encoding a symmetric Heisenberg Hamiltonian. $n$ is the system size, $D$ the dimension of the set $\mathcal{B}_\mathrm{G}$, $M_\mathrm{thres}$ the number of layers and $N_\mathrm{p,c}$ the total number of circuit parameters for the corresponding circuit with $M_\mathrm{thres}$ layers. In each field the first value is attributed to a structured search and the second value is the optimization with random generator sequences. Empty values indicate that the computational cost was too large for an exact optimization.}
        \label{tab:GQSPres}
    \end{table*}
    For all input Hamiltonians the number of required circuit parameters coincides closely with the dimension $D$ of the basis $\mathcal{B}_\mathrm{G}$ supporting the implementability conjecture \eqref{eq:Fspan}. We further validate the conjecture by encoding other matrices from $\mathrm{span}_\mathbb{C}(\mathcal{B}_\mathrm{G})$. They were constructed by choosing random complex coefficients for the linear combination of the $B_i \in \mathcal{B}_\mathrm{G}$ and making the resulting matrix hermitian. We found no case where this method leads to a non-implementable target matrix using the operators generated by $\mathcal{G}$. This was also observed for non-hermitian target matrices and non-hermitian ansatz circuits. Here, the value of $M_\mathrm{thres}$ is twice as large due to the fact that arbitrary non-hermitian matrices have twice the degrees of freedom compared to an arbitrary hermitian matrix.
    
    In terms of optimization strategy, it turns out that \textit{random layering} is as effective in finding the optimal encoding layer as \textit{search algorithms} using a tree search. The \textit{random layering} method was performed by starting at a number of circuit layers that is 5\% higher than it would be required from $N_\mathrm{p,c} = \mathrm{dim}(\mathcal{B}_\mathrm{G})$. We then remove layers until the convergence fails for 10 different random sequences, each optimized for 5 random initial parameter sets.
\section{Benchmark}
\label{section5}
\begin{figure*}[t]
    \centering
    \subfloat[]{\includegraphics{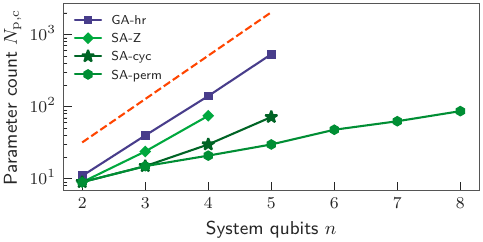}\label{fig:NpHeis}}%
    \hfill 
    \subfloat[]{\includegraphics{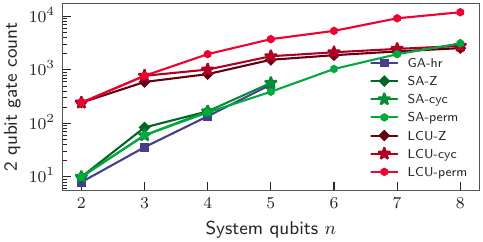}\label{fig:NnlgHeis}}
    \caption{(\textbf{a}): Parameters needed to exactly encode the Heisenberg Hamiltonian using generic (GA)/symmetric (SA) type circuits. GA-hr indicates the hermitian and real generic ansatz. The dashed line marks the lower theoretical bound for parameters in an arbitrary, complex-valued input matrix. (\textbf{b}): Number of 2 qubit gates of the VCBE method and LCU block-encoding of the Heisenberg Hamiltonian for different symmetries.}
\end{figure*}

This section systematically analyzes the resources needed for block-encoding the Heisenberg Hamiltonian from equation \eqref{eq:Heisenberg_H} using methods established in this work.
System sizes up to 8 qubits are considered, as higher qubit numbers are computationally demanding due to a combination of exponentially growing cost for matrix calculations and very high-dimensional optimization landscapes. The minimal number of variational circuit parameters to encode the Heisenberg Hamiltonian for generic circuit ansatze and different symmetric ansatz circuits are shown in Figure \ref{fig:NpHeis}. These values are for exact encoding i.e., the variational circuits have the minimum threshold depth required for reaching $\epsilon \lesssim 10^{-12}$. All circuits are decomposed into single- and two-qubit gates without regard for any underlying architectural restriction\footnote{The code to run VCBE and perform the benchmarks can be made available upon request.}.

 Clearly, all ansatze encode the respective inputs with parameter counts that are lower than for general matrices ($N_\mathrm{p} = 2 \cdot 4^n$). This is due to the symmetry, hermiticity and fully real-valued restrictions on the ansatze. While the parameter count grows exponentially for the circuits respecting $Z_\mathrm{2,xz}$ and $C_\mathrm{n}$ symmetries, it grows polynomially for the circuits with $S_\mathrm{n}$ symmetry.
 
 A comparison of the 2 qubit gate count corresponding to different VCBE circuits is shown in Figure \ref{fig:NnlgHeis}. The number of 2 qubit gates scales similarly to the variational parameter counts. This is a direct result of the $a$-ratio defined in eq. \eqref{eq:aratio}: for a real and hermitian ansatz $a=1$, symmetric ansatze have $a=1/6(n-1)$, $a=1/6n$ and $a=1/3n(n-1)$ for $Z_{2,xz}$,\ $C_\mathrm{n}$ and $S_\mathrm{n}$ symmetries respectively. Note that the permutation invariant circuits retain their polynomial growth in resource requirements.

As a comparison, the resources needed to perform block-encoding using the LCU-method~\cite{Childs2012HamiltonianOperations} are shown in Figure \ref{fig:NnlgHeis}. The \texttt{prepare}-step of the LCU-method needs up to $2^m-2$ CNOT gates to prepare arbitrary $m$-qubit states with real coefficients~\cite{Shende_2006}, depending on the number of terms in the Hamiltonian. Multi-controlled gates in the \texttt{select} unitary are decomposed as in~\cite{10.1109/TCAD.2023.3327102}. It can be seen that VCBE decreases the 2 qubit gate counts, one of the most important quantum resource metrics, by more than an order of magnitude compared to LCU for this specific case of Heisenberg Hamiltonians of up to 5 sites (permutation invariant circuits up to 8 sites) considered in this work. For larger systems, the 2 qubit counts for LCU seem to be smaller than for VCBE, which is expected because, the number of LCU terms grows polynomially, but the size of the VCBE circuits grows exponentially for all ansatze except the permutation invariant ansatz.
\section{Conclusion}
\label{section6}
This work has demonstrated the potential of variational compilation methods for efficiently constructing block-encoding operators, a critical component of quantum signal processing algorithms. We have shown that VCBE can achieve near-optimal parameter counts for encoding arbitrary $2^n \times 2^n$ matrices approaching the theoretical lower bound with generic hardware-efficient circuits. Furthermore, by tailoring the ansatz circuit to match properties of the input matrix, i.e., hermiticity and different symmetries, we achieved a reduction in the number of parameters and enhanced optimizability.

VCBE offers advantages over LCU in terms of gate counts, ancilla count and being viable for $\norm{A/\alpha}_2 \lesssim 1$, but the heavy classical pre-processing associated with optimizing the variational parameters remains the key limitation, restricting the current applicability to systems with $n \le 8$ qubits. Consequently, the most promising near-term application of VCBE may not lie in the direct encoding of large problems. However, VCBE can be used in combination with LCU to reduce the overall resource requirements by only encoding small sub-blocks of the input with VCBE and constructing the full matrix as a linear combination of variational block-encodings.

While in this work we have considered only three symmetries, it remains to be seen how other system-specific properties can be exploited to further reduce the circuit resource requirements. Another interesting aspect is the connection between symmetric VCBE and Multi-variate Quantum Signal Processing~\cite{Rossi2022MultivariableOracle}, potentially leading to methods for determining the ansatz rotation angles without the need for optimizing the complete PQC. Our work also opens a perspective on using generalized QSP-type ansatze as non-unitary ansatze for improving Variational Quantum Eigensolvers and applications like Quantum Machine Learning.
\section*{Acknowledgements}
L.R. ,B.C.M., C.T. and V.M.K. acknowledge funding from the
Ministry of Economic Affairs, Labour and
Tourism Baden-Württemberg in the frame of
the Competence Center Quantum Computing
Baden-Württemberg (project SEQUOIA End-to-End and KQCBW).
S.W. acknowledges funding from the Federal Ministry of Education and
Research (BMBF) under the grants QRydDemo and MUNIQC-Atoms.

\bibliographystyle{quantum}
\bibliography{references}

\begin{thebibliography}{10}

\bibitem{Low2016OptimalProcessing}
Guang~Hao Low and Isaac~L. Chuang.
\newblock ``Optimal hamiltonian simulation by quantum signal processing''.
\newblock \href{https://dx.doi.org/10.1103/PhysRevLett.118.010501}{Phys. Rev. Lett. {\bf 118}, 010501}~(2017).

\bibitem{Low2016HamiltonianQubitization}
Guang~Hao Low and Isaac~L. Chuang.
\newblock ``Hamiltonian {S}imulation by {Q}ubitization''.
\newblock \href{https://dx.doi.org/10.22331/q-2019-07-12-163}{{Quantum} {\bf 3}, 163}~(2019).

\bibitem{Gilyen2018QuantumArithmetics}
Andr\'{a}s Gily\'{e}n, Yuan Su, Guang~Hao Low, and Nathan Wiebe.
\newblock ``Quantum singular value transformation and beyond: exponential improvements for quantum matrix arithmetics''.
\newblock In Proceedings of the 51st Annual ACM SIGACT Symposium on Theory of Computing.
\newblock \href{https://dx.doi.org/10.1145/3313276.3316366}{Page 193–204}.
\newblock STOC 2019. Association for Computing Machinery~(2019).

\bibitem{Motlagh2024}
Danial Motlagh and Nathan Wiebe.
\newblock ``Generalized quantum signal processing''.
\newblock \href{https://dx.doi.org/10.1103/PRXQuantum.5.020368}{PRX Quantum {\bf 5}, 020368}~(2024).

\bibitem{Martyn2021AAlgorithms}
John~M. Martyn, Zane~M. Rossi, Andrew~K. Tan, and Isaac~L. Chuang.
\newblock ``Grand unification of quantum algorithms''.
\newblock \href{https://dx.doi.org/10.1103/PRXQuantum.2.040203}{PRX Quantum {\bf 2}, 040203}~(2021).

\bibitem{Childs2012HamiltonianOperations}
Andrew~M. Childs and Nathan Wiebe.
\newblock ``{Hamiltonian Simulation Using Linear Combinations of Unitary Operations}''.
\newblock \href{https://dx.doi.org/10.26421/QIC12.11-12}{Quantum Information and Computation {\bf 12}, 901–924}~(2012).

\bibitem{Kikuchi2023RealizationComputer}
Yuta Kikuchi, Conor~Mc Keever, Luuk Coopmans, Michael Lubasch, and Marcello Benedetti.
\newblock ``{Realization of quantum signal processing on a noisy quantum computer}''.
\newblock \href{https://dx.doi.org/10.1038/s41534-023-00762-0}{npj Quantum Information {\bf 9}, 93}~(2023).

\bibitem{Madden2022BestProblems}
Liam Madden and Andrea Simonetto.
\newblock ``Best approximate quantum compiling problems''.
\newblock \href{https://dx.doi.org/10.1145/3505181}{ACM Transactions on Quantum Computing {\bf 3}, 1--29}~(2022).

\bibitem{Khatri2018Quantum-assistedCompiling}
Sumeet Khatri, Ryan LaRose, Alexander Poremba, Lukasz Cincio, Andrew~T. Sornborger, and Patrick~J. Coles.
\newblock ``Quantum-assisted quantum compiling''.
\newblock \href{https://dx.doi.org/10.22331/q-2019-05-13-140}{Quantum {\bf 3}, 140}~(2019).

\bibitem{Rakyta_2022}
Péter Rakyta and Zoltán Zimborás.
\newblock ``Approaching the theoretical limit in quantum gate decomposition''.
\newblock \href{https://dx.doi.org/10.22331/q-2022-05-11-710}{Quantum {\bf 6}, 710}~(2022).

\bibitem{Davis2020TowardsSynthesis}
Marc~G. Davis, Ethan Smith, Ana Tudor, Koushik Sen, Irfan Siddiqi, and Costin Iancu.
\newblock ``Towards optimal topology aware quantum circuit synthesis''.
\newblock In 2020 IEEE International Conference on Quantum Computing and Engineering (QCE).
\newblock \href{https://dx.doi.org/10.1109/QCE49297.2020.00036}{Pages 223--234}.
\newblock ~(2020).

\bibitem{Larocca_2025}
Martín Larocca, Supanut Thanasilp, Samson Wang, Kunal Sharma, Jacob Biamonte, Patrick~J. Coles, Lukasz Cincio, Jarrod~R. McClean, Zoë Holmes, and M.~Cerezo.
\newblock ``Barren plateaus in variational quantum computing''.
\newblock \href{https://dx.doi.org/10.1038/s42254-025-00813-9}{Nature Reviews Physics {\bf 7}, 174–189}~(2025).

\bibitem{jax2018github}
James Bradbury, Roy Frostig, Peter Hawkins, Matthew~James Johnson, Chris Leary, Dougal Maclaurin, George Necula, Adam Paszke, Jake Vander{P}las, Skye Wanderman-{M}ilne, and Qiao Zhang~(2018).
\newblock  url:~\url{http://github.com/jax-ml/jax}.

\bibitem{blondel2022efficientmodularimplicitdifferentiation}
Mathieu Blondel, Quentin Berthet, Marco Cuturi, Roy Frostig, Stephan Hoyer, Felipe Llinares-López, Fabian Pedregosa, and Jean-Philippe Vert.
\newblock ``Efficient and modular implicit differentiation''~(2022).
\newblock  \href{http://arxiv.org/abs/2105.15183}{arXiv:2105.15183}.

\bibitem{Shende2004MinimalCircuits}
Vivek~V. Shende, Igor~L. Markov, and Stephen~S. Bullock.
\newblock ``Minimal universal two-qubit controlled-not-based circuits''.
\newblock \href{https://dx.doi.org/10.1103/PhysRevA.69.062321}{Phys. Rev. A {\bf 69}, 062321}~(2004).

\bibitem{Gard_2020}
Bryan~T. Gard, Linghua Zhu, George~S. Barron, Nicholas~J. Mayhall, Sophia~E. Economou, and Edwin Barnes.
\newblock ``Efficient symmetry-preserving state preparation circuits for the variational quantum eigensolver algorithm''.
\newblock \href{https://dx.doi.org/10.1038/s41534-019-0240-1}{npj Quantum Information {\bf 6}, 10}~(2020).

\bibitem{Lacroix_2023}
Denis Lacroix, Edgar~Andres Ruiz~Guzman, and Pooja Siwach.
\newblock ``Symmetry breaking/symmetry preserving circuits and symmetry restoration on quantum computers: A quantum many-body perspective''.
\newblock \href{https://dx.doi.org/10.1140/epja/s10050-022-00911-7}{The European Physical Journal A {\bf 59}, 3}~(2023).

\bibitem{Meyer_2023}
Johannes~Jakob Meyer, Marian Mularski, Elies Gil-Fuster, Antonio~Anna Mele, Francesco Arzani, Alissa Wilms, and Jens Eisert.
\newblock ``Exploiting symmetry in variational quantum machine learning''.
\newblock \href{https://dx.doi.org/10.1103/PRXQuantum.4.010328}{PRX Quantum {\bf 4}, 010328}~(2023).

\bibitem{Larocca_2022}
Martín Larocca, Frédéric Sauvage, Faris~M. Sbahi, Guillaume Verdon, Patrick~J. Coles, and M.~Cerezo.
\newblock ``Group-invariant quantum machine learning''.
\newblock \href{https://dx.doi.org/10.1103/prxquantum.3.030341}{PRX Quantum{\bf 3}}~(2022).

\bibitem{Mansky_2025}
Maximilian~Balthasar Mansky, Miguel~Armayor Martinez, Alejandro~Bravo de~la Serna, Santiago Londoño~Castillo, Dimitra Nikolaidou, Gautham Sathish, Zhihao Wang, Sebastian Wölckert, and Claudia Linnhoff-Popien.
\newblock ``Scaling of symmetry-restricted lie groups''.
\newblock \href{https://dx.doi.org/10.1088/1402-4896/adf0ee}{Physica Scripta {\bf 100}, 085222}~(2025).

\bibitem{younis2020qfastquantumsynthesisusing}
Ed~Younis, Koushik Sen, Katherine Yelick, and Costin Iancu.
\newblock ``Qfast: Quantum synthesis using a hierarchical continuous circuit space''~(2020).
\newblock  \href{http://arxiv.org/abs/2003.04462}{arXiv:2003.04462}.

\bibitem{Shende_2006}
V.V. Shende, S.S. Bullock, and I.L. Markov.
\newblock ``Synthesis of quantum-logic circuits''.
\newblock \href{https://dx.doi.org/10.1109/tcad.2005.855930}{IEEE Transactions on Computer-Aided Design of Integrated Circuits and Systems {\bf 25}, 1000–1010}~(2006).

\bibitem{10.1109/TCAD.2023.3327102}
Rafaella Vale, Thiago Melo~D. Azevedo, Ismael C.~S. Ara\'{u}jo, Israel~F. Araujo, and Adenilton~J. da~Silva.
\newblock ``Circuit decomposition of multicontrolled special unitary single-qubit gates''.
\newblock \href{https://dx.doi.org/10.1109/TCAD.2023.3327102}{Trans. Comp.-Aided Des. Integ. Cir. Sys. {\bf 43}, 802–811}~(2024).

\bibitem{Rossi2022MultivariableOracle}
Zane~M. Rossi and Isaac~L. Chuang.
\newblock ``{Multivariable quantum signal processing (M-QSP): prophecies of the two-headed oracle}''.
\newblock \href{https://dx.doi.org/10.22331/q-2022-09-20-811}{Quantum {\bf 6}, 811}~(2022).

\bibitem{Nemeth2023OnCharacterizations}
Balázs Németh, Blanka Kövér, Boglárka Kulcsár, Roland~Botond Miklósi, and András Gilyén.
\newblock ``On variants of multivariate quantum signal processing and their characterizations''~(2023).
\newblock  \href{http://arxiv.org/abs/2312.09072}{arXiv:2312.09072}.

\bibitem{Wiersema_2024}
Roeland Wiersema, Efekan Kökcü, Alexander~F. Kemper, and Bojko~N. Bakalov.
\newblock ``Classification of dynamical lie algebras of 2-local spin systems on linear, circular and fully connected topologies''.
\newblock \href{https://dx.doi.org/10.1038/s41534-024-00900-2}{npj Quantum Information {\bf 10}, 110}~(2024).

\bibitem{allcock2024dynamicalliealgebrasquantum}
Jonathan Allcock, Miklos Santha, Pei Yuan, and Shengyu Zhang.
\newblock ``On the dynamical lie algebras of quantum approximate optimization algorithms''~(2024).
\newblock  \href{http://arxiv.org/abs/2407.12587}{arXiv:2407.12587}.

\end{thebibliography}

\onecolumn

\appendix

    
\section{Additional generic circuits and their optimization characteristics}
\label{app:AddGenCircs}
The main text only evaluates a single type of generic ansatz circuit which leaves open the reliability of VCBE over a range of different types of generic ansatz circuits. An additional 15 circuits are evaluated in this section consolidating and complementing the empirical evidence for a stable optimization near the theoretical lower bound.
    
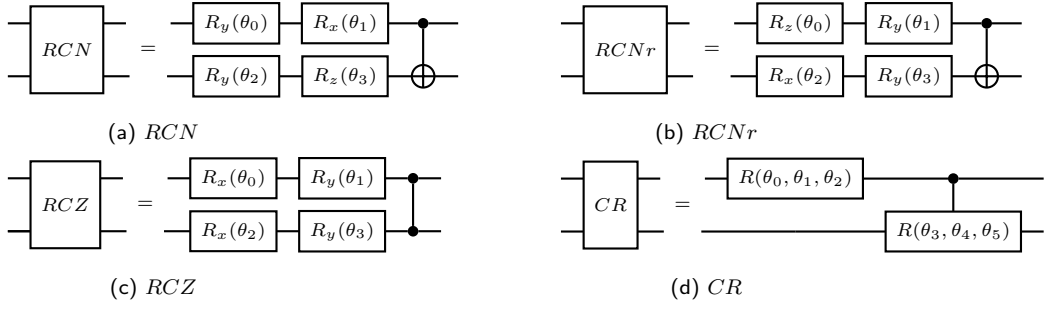
\begin{figure}[ht]
        \centering
        \hspace{-2cm}
        \begin{subfigure}{0.24\textwidth}
            \centering
            \begin{quantikz}[row sep = {0.7cm,between origins}, column sep = 0.3cm,font=\scriptsize]
                & \gate[2]{RCN} & \qw \midstick[2,brackets=none]{=} & \gate{R_y(\theta_0)} & \gate{R_x(\theta_1)} & \ctrl{1} & \qw\\
                & & \qw & \gate{R_y(\theta_2)} & \gate{R_z(\theta_3)} & \targ{} & \qw
            \end{quantikz}
            \caption{$RCN$}
        \end{subfigure}
        \hspace{3cm}
        \begin{subfigure}{0.24\textwidth}
            \centering
            \begin{quantikz}[row sep = {0.7cm,between origins}, column sep = 0.3cm,font=\scriptsize]
                & \gate[2]{RCNr} & \qw \midstick[2,brackets=none]{=} & \gate{R_z(\theta_0)} & \gate{R_y(\theta_1)} & \ctrl{1} & \qw\\
                & & \qw & \gate{R_x(\theta_2)} & \gate{R_y(\theta_3)} & \targ{} & \qw
            \end{quantikz}
            \caption{$RCNr$}
        \end{subfigure}

        \hspace{-2cm}
        \begin{subfigure}{0.24\textwidth}
            \centering
            \begin{quantikz}[row sep = {0.7cm,between origins}, column sep = 0.3cm,font=\scriptsize]
                & \gate[2]{RCZ} & \qw \midstick[2,brackets=none]{=} & \gate{R_x(\theta_0)} & \gate{R_y(\theta_1)} & \ctrl{1} & \qw\\
                & \qw & \qw & \gate{R_x(\theta_2)} & \gate{R_y(\theta_3)} & \control{} & \qw
            \end{quantikz}
            \caption{$RCZ$}
        \end{subfigure}
        \hspace{3cm}
        \begin{subfigure}{0.24\textwidth}
            \centering
            \begin{quantikz}[row sep = {0.7cm,between origins}, column sep = 0.3cm,font=\scriptsize]
                & \gate[2]{CR} & \qw \midstick[2,brackets=none]{=} & \gate{R(\theta_0,\theta_1,\theta_2)} & \ctrl{1} & \qw\\
                & & \qw & \qw & \gate{R(\theta_3,\theta_4,\theta_5)} & \qw
            \end{quantikz}
            \caption{$CR$}
        \end{subfigure}
        
        \caption{2 qubit circuit blocks used in the generic ansatze. $RCN$ and $RCNr$ gate are CNOT-based circuits with an optimal number of parameters per CNOT when inserted into larger circuits. They use different rotation gates. $RCZ$ is CZ-based. $CR$ uses a general controlled rotation and is not optimal in a wider circuit, since the Z-rotation of the single qubit gate can be passed through the control.}
        \label{fig:2qblocks}
    \end{figure}

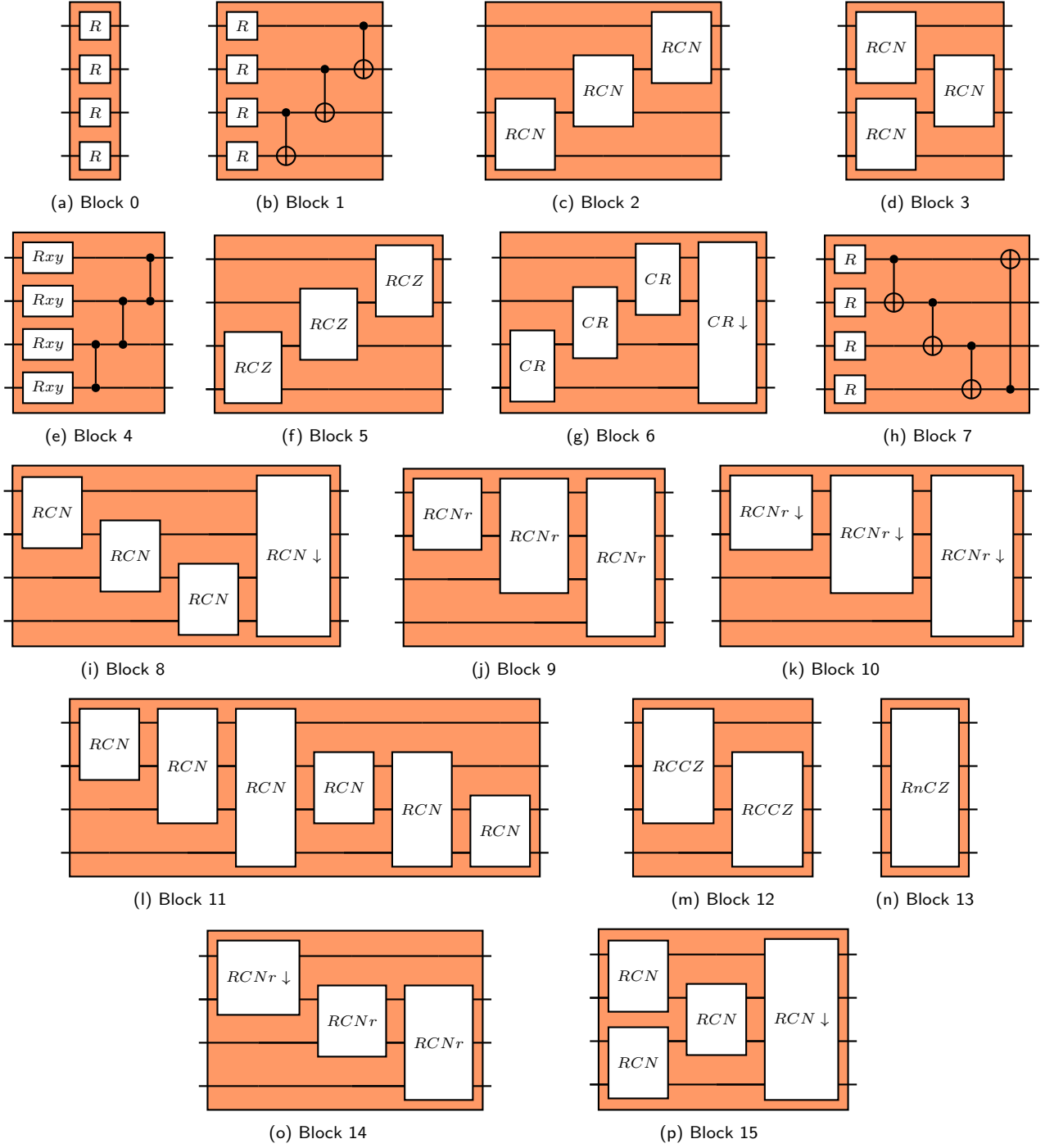
\begin{figure}
        \hspace{-0.6cm}
        \begin{subfigure}{0.24\textwidth}
            \centering
            \begin{quantikz}[row sep = {0.7cm,between origins}, column sep = 0.3cm,font=\scriptsize]
                & \gate{R}\gategroup[4,steps=1,style={inner sep=1pt,fill=atomictangerine},background]{} & \qw\\
                & \gate{R} & \qw\\
                & \gate{R} & \qw\\
                & \gate{R} & \qw
            \end{quantikz}
            \caption{Block 0}
        \end{subfigure}
        \hspace{-1cm}
        \begin{subfigure}[b]{0.24\textwidth}
            \centering
            \begin{quantikz}[row sep = {0.7cm,between origins}, column sep = 0.3cm,font=\scriptsize]
                & \gate{R}\gategroup[4,steps=4,style={inner sep=1pt,fill=atomictangerine},background]{} & \qw & \qw & \ctrl{1} & \qw\\
                & \gate{R} & \qw & \ctrl{1} & \targ{} & \qw\\
                & \gate{R} & \ctrl{1} & \targ{} & \qw & \qw\\
                & \gate{R} & \targ{} & \qw & \qw & \qw
            \end{quantikz}
            \caption{Block 1}
        \end{subfigure}
        \hfill
        \begin{subfigure}[b]{0.24\textwidth}
            \centering
            \begin{quantikz}[row sep = {0.7cm,between origins}, column sep = 0.3cm,font=\scriptsize]
                & \qw\gategroup[4,steps=3,style={inner sep=1pt,fill=atomictangerine},background]{} & \qw & \gate[2]{RCN} & \qw\\
                & \qw & \gate[2]{RCN} & \qw & \qw\\
                & \gate[2]{RCN} & \qw & \qw & \qw\\
                & \qw & \qw & \qw & \qw
            \end{quantikz}
            \caption{Block 2}
        \end{subfigure}
        \hspace{1cm}
        \begin{subfigure}[b]{0.24\textwidth}
            \centering
            \begin{quantikz}[row sep = {0.7cm,between origins}, column sep = 0.3cm,font=\scriptsize]
                & \gate[2]{RCN}\gategroup[4,steps=2,style={inner sep=1pt,fill=atomictangerine},background]{} & \qw & \qw\\
                & \qw & \gate[2]{RCN} & \qw\\
                & \gate[2]{RCN} & \qw & \qw\\
                & \qw & \qw & \qw
            \end{quantikz}
            \caption{Block 3}
        \end{subfigure}

        \hspace{-0.7cm}
        \begin{subfigure}[b]{0.24\textwidth}
            \centering
            \begin{quantikz}[row sep = {0.7cm,between origins}, column sep = 0.3cm,font=\scriptsize]
                & \gate{Rxy}\gategroup[4,steps=4,style={inner sep=1pt,fill=atomictangerine},background]{} & \qw & \qw & \ctrl{1} & \qw\\
                & \gate{Rxy} & \qw & \ctrl{1} & \control{} & \qw \\
                & \gate{Rxy} & \ctrl{1} & \control{} & \qw & \qw\\
                & \gate{Rxy} & \control{} & \qw & \qw & \qw
            \end{quantikz}
            \caption{Block 4}
        \end{subfigure}
        \hspace{-0.5cm}
        \begin{subfigure}[b]{0.24\textwidth}
            \centering
            \begin{quantikz}[row sep = {0.7cm,between origins}, column sep = 0.3cm,font=\scriptsize]
                & \qw\gategroup[4,steps=3,style={inner sep=1pt,fill=atomictangerine},background]{} & \qw & \gate[2]{RCZ} & \qw\\
                & \qw & \gate[2]{RCZ} & \qw & \qw\\
                & \gate[2]{RCZ} & \qw & \qw & \qw\\
                & \qw & \qw & \qw & \qw
            \end{quantikz}
            \caption{Block 5}
        \end{subfigure}
        \hspace{0.3cm}
        \begin{subfigure}[b]{0.24\textwidth}
            \centering
            \begin{quantikz}[row sep = {0.7cm,between origins}, column sep = 0.3cm,font=\scriptsize]
                & \qw\gategroup[4,steps=4,style={inner sep=1pt,fill=atomictangerine},background]{} & \qw & \gate[2]{CR} & \gate[4]{CR\downarrow} & \qw\\
                & \qw & \gate[2]{CR} & \qw & \qw & \qw\\
                & \gate[2]{CR} & \qw & \qw & \qw & \qw\\
                & \qw & \qw & \qw & \qw & \qw
            \end{quantikz}
            \caption{Block 6}
        \end{subfigure}
        \hspace{0.8cm}
        \begin{subfigure}[b]{0.24\textwidth}
            \centering
            \begin{quantikz}[row sep = {0.7cm,between origins}, column sep = 0.3cm,font=\scriptsize]
                & \gate{R}\gategroup[4,steps=5,style={inner sep=1pt,fill=atomictangerine},background]{} & \ctrl{1} & \qw & \qw & \targ{} & \qw\\
                & \gate{R} & \targ{} & \ctrl{1} & \qw & \qw & \qw\\
                & \gate{R} & \qw & \targ{} & \ctrl{1} & \qw & \qw\\
                & \gate{R} & \qw & \qw & \targ{} & \ctrl{-3} & \qw
            \end{quantikz}
            \caption{Block 7}
        \end{subfigure}
        \hspace{-0.5cm}
    
        \hspace{-0.15cm}
        \begin{subfigure}[b]{0.24\textwidth}
            \centering
            \begin{quantikz}[row sep = {0.7cm,between origins}, column sep = 0.3cm,font=\scriptsize]
                & \gate[2]{RCN}\gategroup[4,steps=4,style={inner sep=1pt,fill=atomictangerine},background]{} & \qw & \qw & \gate[4]{RCN\downarrow} & \qw\\
                & \qw & \gate[2]{RCN} & \qw & \qw & \qw\\
                & \qw & \qw & \gate[2]{RCN} & \qw & \qw\\
                & \qw & \qw & \qw & \qw & \qw
            \end{quantikz}
            \caption{Block 8}
        \end{subfigure}
        \hspace{2cm}
        \begin{subfigure}[b]{0.24\textwidth}
            \centering
            \begin{quantikz}[row sep = {0.7cm,between origins}, column sep = 0.3cm,font=\scriptsize]
                & \gate[2]{RCNr}\gategroup[4,steps=3,style={inner sep=1pt,fill=atomictangerine},background]{} & \gate[3]{RCNr} & \gate[4]{RCNr} & \qw\\
                & \qw & \qw & \qw & \qw\\
                & \qw & \qw & \qw & \qw\\
                & \qw & \qw & \qw & \qw
            \end{quantikz}
            \caption{Block 9}
        \end{subfigure}
        \hspace{0.8cm}
        \begin{subfigure}[b]{0.24\textwidth}
            \centering
            \begin{quantikz}[row sep = {0.7cm,between origins}, column sep = 0.3cm,font=\scriptsize]
                & \gate[2]{RCNr\downarrow}\gategroup[4,steps=3,style={inner sep=1pt,fill=atomictangerine},background]{} & \gate[3]{RCNr\downarrow} & \gate[4]{RCNr\downarrow} & \qw\\
                & \qw & \qw & \qw & \qw\\
                & \qw & \qw & \qw & \qw\\
                & \qw & \qw & \qw & \qw
            \end{quantikz}
            \caption{Block 10}
        \end{subfigure}
        \hspace{-0.5cm}
    
        \hspace{0.8cm}
        \begin{subfigure}[b]{0.24\textwidth}
            \centering
            \begin{quantikz}[row sep = {0.7cm,between origins}, column sep = 0.3cm,font=\scriptsize]
                & \gate[2]{RCN}\gategroup[4,steps=6,style={inner sep=1pt,fill=atomictangerine},background]{} & \gate[3]{RCN} & \gate[4]{RCN} & \qw & \qw & \qw & \qw\\
                & \qw & \qw & \qw & \gate[2]{RCN} & \gate[3]{RCN} & \qw & \qw\\
                & \qw & \qw & \qw & \qw & \qw & \gate[2]{RCN} & \qw\\
                & \qw & \qw & \qw & \qw & \qw & \qw & \qw 
            \end{quantikz}
            \caption{Block 11}
        \end{subfigure}
        \hspace{4.5cm}
        \begin{subfigure}[b]{0.24\textwidth}
            \centering
            \begin{quantikz}[row sep = {0.7cm,between origins}, column sep = 0.3cm,font=\scriptsize]
                & \gate[3]{RCCZ}\gategroup[4,steps=2,style={inner sep=1pt,fill=atomictangerine},background]{} & \qw & \qw\\
                & \qw & \gate[3]{RCCZ} & \qw\\
                & \qw & \qw & \qw\\
                & \qw & \qw & \qw
            \end{quantikz}
            \caption{Block 12}
        \end{subfigure}
        \hspace{-1cm}
        \begin{subfigure}[b]{0.24\textwidth}
            \centering
            \begin{quantikz}[row sep = {0.7cm,between origins}, column sep = 0.3cm,font=\scriptsize]
                & \gate[4]{RnCZ}\gategroup[4,steps=1,style={inner sep=1pt,fill=atomictangerine},background]{} & \qw\\
                & \ghost{H} & \qw\\
                & \ghost{H} & \qw\\
                & \ghost{H} & \qw
            \end{quantikz}
            \caption{Block 13}
        \end{subfigure}

        \hspace{3cm}
        \begin{subfigure}[b]{0.24\textwidth}
            \centering
            \begin{quantikz}[row sep = {0.7cm,between origins}, column sep = 0.3cm,font=\scriptsize]
                & \gate[2]{RCNr\downarrow}\gategroup[4,steps=3,style={inner sep=1pt,fill=atomictangerine},background]{} & \qw & \qw & \qw\\
                & \qw & \gate[2]{RCNr} & \gate[3]{RCNr}  & \qw\\
                & \qw & \qw & \qw & \qw\\
                & \qw & \qw & \qw & \qw
            \end{quantikz}
            \caption{Block 14}
        \end{subfigure}
        \hspace{2cm}
        \begin{subfigure}[b]{0.24\textwidth}
            \centering
            \begin{quantikz}[row sep = {0.7cm,between origins}, column sep = 0.3cm,font=\scriptsize]
                & \gate[2]{RCN}\gategroup[4,steps=3,style={inner sep=1pt,fill=atomictangerine},background]{} & \qw & \gate[4]{RCN\downarrow} & \qw\\
                & \qw & \gate[2]{RCN} & \qw  & \qw\\
                & \gate[2]{RCN} & \qw & \qw & \qw\\
                & \qw & \qw & \qw & \qw
            \end{quantikz}
            \caption{Block 15}
        \end{subfigure}
    
        \caption{$N=4$ Qubit example circuit blocks. The downward-facing arrow on some gates indicates the direction of the CNOT gate. All individual R, RCN(r) Rxy and R(C/n)CZ in a circuits have independent parameters.}
        \label{fig:quantum_circuits}
    \end{figure}

Quantum computers have widely varying underlying architectures in regard to gate sets and connectivity. The generic ansatz used in the main text for example, can be easily used on hardware using CNOT gates (up to local single qubit rotations) and a next-nearest neighbor coupling. Different configurations and multi-qubit gates will now be explored systematically.
The basic 2 qubit building blocks used in the additional generic ansatze are presented in Figure \ref{fig:2qblocks}.
Also, $RCCZ$ and $RnCZ$ gates are defined, that are 3  and $n$ qubit $RCZ$ gates. Real versions of all gates are extracted by setting the angles $\theta$ of $R_\mathrm{x}$  and $R_\mathrm{z}$ gates to $\theta = 0$ and general rotations $R(\theta_0,\theta_1,\theta_2)$ to $R(\theta_0,0,0)$.
Combining these basic gate constructs, we define 16 unique ansatz blocks as listed in Figure \ref{fig:quantum_circuits}.


Each circuit has unique specifications (the parameter to gate ratio $a$ is defined in eq. \eqref{eq:aratio}):
    \begin{itemize}
        \setlength{\itemsep}{0pt}
        \item Block 0: No entanglement
        \item Block 1: linear entanglement, not optimal parameter per non-local gate ratio $a$
        \item Block 2: linear entanglement, optimal $a$, CNOT gates
        \item Block 3: linear entanglement, optimal $a$, CNOT gates, optimal depth circuit through parallel $RCN$ gates
        \item Block 4: linear entanglement, not optimal $a$, CZ gates
        \item Block 5: linear entanglement, optimal $a$, CZ gates
        \item Block 6: circular entanglement, not optimal $a$, CR gates
        \item Block 7: circular entanglement, not optimal $a$, CNOT gates
        \item Block 8: circular entanglement, optimal $a$, CNOT gates
        \item Block 9: star configuration, optimal $a$, CNOT gates, CNOT control is on ancilla
        \item Block 10: star configuration, optimal $a$, CNOT gates, CNOT target is on ancilla
        \item Block 11: all-to-all configuration, optimal $a$, CNOT gates
        \item Block 12: linear CCZ, optimal $a$
        \item Block 13: full nCZ, optimal $a$
        \item Block 14: star configuration, optimal $a$, CNOT control is on system qubit
        \item Block 15: circular entanglement, optimal $a$, CNOT gates, optimal depth circuit
    \end{itemize}
    
\begin{figure*}[t]
        \centering
        \subfloat[]{\includegraphics[width=0.48\textwidth]{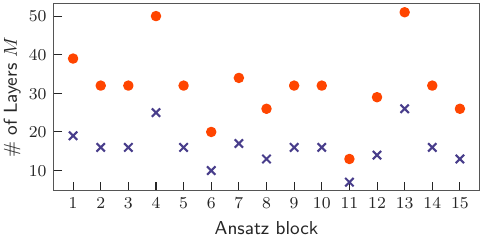}}\hfill
        \subfloat[]{\includegraphics[width=0.48\textwidth]{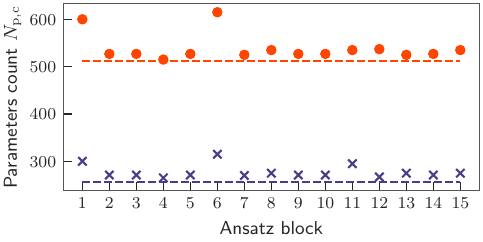}}\\[2mm]
        \subfloat[]{\includegraphics[width=0.48\textwidth]{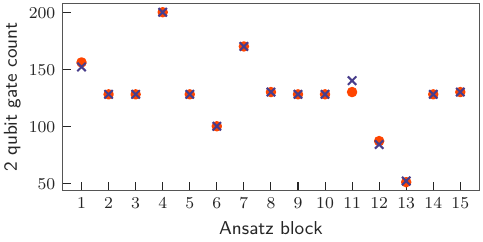}}\hfill
        \subfloat[]{\includegraphics[width=0.48\textwidth]{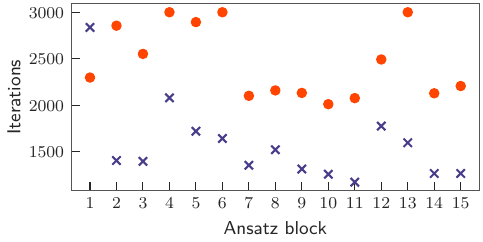}}
        \caption{\textbf{(a):} Number of circuit layers, \textbf{(b):} variational parameters and \textbf{(c):} non-local gates (NlGs) needed for an exact optimization of a random $2^4 \times 2^4$ input matrix using a single ancilla qubit and 4 system qubits. \textbf{(d):} Number of iterations were taken from the best of runs with 10 random matrices and 10 random initializations each.\\
        Blue dots indicate results from arbitrary complex square matrices and green crosses are results from arbitrary hermitian inputs with hermitian ansatz circuits. Error bars intdicating the variance of the 100 runs in total are omitted since optimization runs with different random inputs and random initialization lead to practically identical performance. Dashed blue and green lines are the theoretical lower bounds for parameter numbers at 256 and 512. Optimizer iterations are capped at 3000.}
        \label{fig:4S_res}
    \end{figure*}

    Figures \ref{fig:4S_res} show the performance of an exact compilation of each of the ansatz blocks. Here, we have only given the results for a 4 qubit input matrix and 1 ancilla qubit. All other system sizes ($n=2,3,5$) and possible input/circuit configurations (real, (non)-hermitian) show almost identical and predictable behaviors that have already been discussed in the main text.
    
    A few important remarks can now be listed:
    \begin{enumerate}
        \item Block 0 fails to optimize since no entanglement between ancilla and system qubit means that no non-unitary matrices can be encoded.
        \item Circuit configuration is practically irrelevant for the success of encoding the input exactly. Linear, circular, star, and all-to-all architectures lead to identical results in terms of circuit parameter counts. Even if the ancilla qubit is not the main hub (see ansatz block 14), the ansatz optimizes.
        \item The same applies to the gate sets used. Success of optimization is only reliant on the amount of variational parameters in the circuit. 2-qubit gates work as well as $n$-qubit gates, as long as the inter-connectivity is sufficient.
        \item A divergence from the theoretical lower bound in parameters comes from an overhead from adding full layers with multiple parameters and not optimal placement of variational gates that lead to cancellations as seen in blocks 1 and 6 (example is the $R_\mathrm{z}$ gates that can be pushed through the control of a CNOT).
        In case the circuit layout does not have an optimal ratio, free parameters of the input do not get absorbed by the additional parameterized gates, but rather the circuit depths increases by the overhead in the ratio. 
        \item The expected exponential growth for larger system sizes affects all ansatz blocks likewise.
    \end{enumerate}
    The listed observations have also been described in unitary compilation \cite{Madden2022BestProblems}.\\
    
    \subsection{Multiple ancilla qubits}
    \label{app:ancN}
    Up to now, the ansatz circuits contained only a single ancilla qubit, which is the simplest block-encoding scheme. Standard block-encoding techniques usually include multiple ancillas, depending on input structure and sparsity. Table \ref{tab:A_n_4S} indicates that for a $4+m$ qubit ansatz, a trend towards higher gate counts in variational and two-qubit gates for more ancillas exists. This means that independent input parameters cannot be shifted into the ancilla register freely as a method to trade in circuit depth for circuit width.
    \begin{table}[h]
        \centering
        \begin{tabular}{c|ccc}
            Ancillas & 1 & 2 & 3\\\hline
            Parameter count & 141 (136) & 156 (136) & 175 (136)\\
            NlG count & 136 (66) & 156 (66) & 168 (66)\\
        \end{tabular}
        \caption{Resources to exactly encode a $2^4 \times 2^4$ complex, hermitian matrix with $m=1,2,3$ BE ancillas using the generic circuit block from Section \ref{sec:standlayan}. Optimal resource counts are given in brackets.}
        \label{tab:A_n_4S}
    \end{table}
    This clearly refutes the better trainability of the ansatz with higher qubit counts which Kikuchi et al. \cite{Kikuchi2023RealizationComputer} found. The reason could be that their ansatz does not have an optimal $a$-ratio which a wider circuit could make up for.

 

\section{Pauli Gadgets}
\label{app:GQSPcircuits}
    The symmetric ansatz circuits consist of pauli gadgets if the generators allow for this decomposition. Figure \ref{fig:PauliGadget} gives an example circuit for a pauli gadget implementing the operator
    \begin{equation}
        P(\theta) = \mathrm{e}^{\theta i (ZZI + ZIZ + IZZ)},
    \end{equation}
    which is present in permutation and cyclic invariant generator sets for a 3 qubit system. It can be decomposed as
    \begin{equation}
        P(\theta) = \mathrm{e}^{\theta i ZZI}\mathrm{e}^{\theta i ZIZ}\mathrm{e}^{\theta i IZZ},
    \end{equation}
    since $ZZI$, $ZIZ$ and $IZZ$ commute. The ansatz requires a controlled version of $P(\theta)$ and the corresponding circuit is drawn in Figure \ref{fig:cP}, where only the rotation gates are controlled. Each can be implemented using two CNOT gates, or if the specific two-qubit gate is native to a hardware.
    
    \begin{figure}[ht]
        \centering
        \begin{quantikz}[row sep = {0.7cm,between origins}, column sep = 0.3cm,font=\small]
            & \gate[3]{P} & \qw \midstick[3,brackets=none]{=} & \targ{} & \gate{R_\mathrm{z}(\theta)} & \targ{} & \qw & \qw & \qw & \targ{} & \gate{R_\mathrm{z}(\theta)} & \targ{} & \qw\\
            & \qw & \qw & \ctrl{-1} & \qw & \ctrl{-1} & \targ{} & \gate{R_\mathrm{z}(\theta)} & \targ{} & \qw & \qw & \qw & \qw\\
            & \qw & \qw & \qw & \qw & \qw & \ctrl{-1} & \qw & \ctrl{-1} & \ctrl{-2} & \qw & \ctrl{-2} & \qw
        \end{quantikz}
        \caption{Circuit for Pauli gadget.}
        \label{fig:PauliGadget}
    \end{figure}
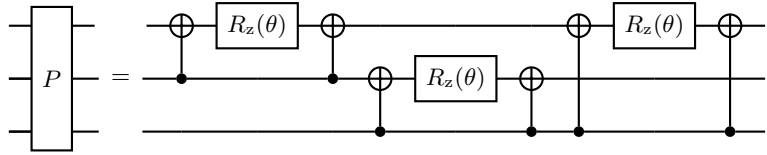
    \begin{figure}[th]
        \centering
        \begin{quantikz}[row sep = {0.7cm,between origins}, column sep = 0.3cm,font=\small]
            & \ctrl{1} & \qw \midstick[4,brackets=none]{=} & \qw & \ctrl{1} & \qw & \qw & \ctrl{2} & \qw & \qw & \ctrl{1} & \qw\\
            & \gate[3]{cP} & \qw & \targ{} & \gate{R_\mathrm{z}(\theta)} & \targ{} & \qw & \qw & \qw & \targ{} & \gate{R_\mathrm{z}(\theta)} & \targ{} & \qw\\
            & \qw & \qw & \ctrl{-1} & \qw & \ctrl{-1} & \targ{} & \gate{R_\mathrm{z}(\theta)} & \targ{} & \qw & \qw & \qw & \qw\\
            & \qw & \qw & \qw & \qw & \qw & \ctrl{-1} & \qw & \ctrl{-1} & \ctrl{-2} & \qw & \ctrl{-2} & \qw
        \end{quantikz}
        \caption{Circuit for controlled Pauli gadget.}
        \label{fig:cP}
    \end{figure}
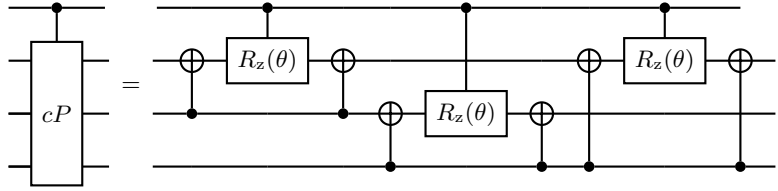

    The connectivity required in the quantum architecture on the system qubits is specific to the input symmetry. A Hamiltonian of a chain needs only linear nearest-neighbor connectivity, whereas a permutation invariant system requires an all-to-all connectivity. The single ancilla has to be linked to all qubits in general. Explicit circuits can lift this requirement as can be seen for Pauli gadget circuits.

            
            

\section{Controlled VCBE circuits}
    It is often necessary to construct controlled versions of block-encoding circuits for an embedding into larger algorithms. Standard layered circuits are complex add a control, since controlled generic circuits are produced by controlling all gates and a controlled version of standard GQSP-type ansatze is done by controlling the single qubit ancilla rotations and additionally the $cP_{i}$ operators turning them into $ccP_{i}$ gates.\\

\begin{figure}[ht]
        \centering
        \begin{quantikz}[row sep = {0.7cm,between origins}, column sep = 0.3cm,font=\small]
            & \qw & \ctrl{1} & \qw \midstick[4,brackets=none]{=} & \qw & \ctrl{1} & \qw & \qw\\
            & \qwbundle{m} & \gate[2,style={fill=brightturquoise}]{U_\mathrm{var,h}(\bm{\theta})} & \qw  & \gate[2,style={fill=blue!10}]{U_\mathrm{var}(\bm{\theta})^\dagger} & \gate[2,style={fill=darkgoldenrod}]{V} & \gate[2,style={fill=blue!10}]{U_\mathrm{var}(\bm{\theta})} & \qw \\
            & \qwbundle{n} & \qw & \qw & \qw & \qw  & \qw & \qw 
        \end{quantikz}
        \caption{Controlled version of a hermitian block encoding ansatz.}
        \label{fig:cherm}
    \end{figure}
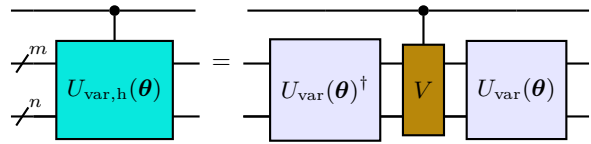

    Using the hermitian circuit structure can significantly simplify this by requiring only to use a controlled version of $V$ as
    \begin{equation}
        U_\mathrm{var,h}(\bm{\theta}) = (\mathds{1} \otimes U_\mathrm{var}(\bm{\theta})) cV (\mathds{1} \otimes U_\mathrm{var}(\bm{\theta})^\dagger).
    \end{equation}
    The circuit is depicted in Figure \ref{fig:cherm}. In this work $V$ is either hadamard gates on all qubits (generic) or a single hadamard on the ancilla (GQSP-type), making the decomposition of $cV$ trivial.

\section{The Variational Matrix Function and its Generators}
\label{app:Bcalc}
This section derives a method to calculate the set $\mathcal{B}$ of linearly independent operators that construct the matrix function $F$ in the top left block of a GQSP-type variational circuit with generators $\mathcal{G}$. The underlying question is if we can make statements about the expressibility of $F$ from the sequence of generators used. We will outline a general approach to answer this question, but this paper can only conjecture a possibility instead of finding a sound mathematical proof. 
As stated in the main text, the matrix function of the variational ansatz block is
    \begin{equation}
        F = (\bra{0} \otimes \mathds{1}^{\otimes n}) U_\mathrm{var} (\ket{0} \otimes \mathds{1}^{\otimes n})
    \end{equation}
    which evaluates to
    \begin{equation}
        F = \prod_{i=1}^{M} \left ( a_{i} \mathds{1}^{\otimes n} + b_{i} P_{i}(\theta_{i}) \right ) \mathrm{e}^{i\lambda_0},
        \label{eq:Fprod_app}
    \end{equation}
    where the coefficients $a_{i}$, $b_{i}$ and $\lambda_0$ are determined by the single qubit rotations.
    
    If all $P_{i}$ are identical then the sequence boils down to generalized quantum signal processing and the angles can be readily calculated to fit any matrix function $F(P)$ within the restrictions given in \cite{Motlagh2024}. In the more general case of non-identical $P_{i}$, multi-variate QSP \cite{Rossi2022MultivariableOracle,Nemeth2023OnCharacterizations} could be employed to calculate the rotation angles to produce a specific output function $F$. Unfortunately, this can only be done for very specific cases and is one of the largest open questions in the research of QSP-type quantum algorithms.
    
    Rewriting equation \eqref{eq:Fprod_app} as
    \begin{equation}
        F = \sum_{\alpha} \Tilde{a}_{\alpha} P^\alpha,
        \label{eq:FsumP}
    \end{equation}
    with $\alpha = (\alpha_1,\alpha_2,...,\alpha_\mathrm{M})$, $\alpha_{i} \in \{0,1\}$ and $P^\alpha = P_\mathrm{M}^{\alpha_\mathrm{M}} \cdots P_2^{\alpha_2}P_1^{\alpha_1}$ constructed by the generating set $\mathcal{G}_\alpha \subseteq \mathcal{G}$ that are present in each term, we can turn the product representation into a sum.\\
    Utilizing the notion of dynamical Lie algebras(DLA), all terms $P^\alpha$ in the sum can be characterized by their corresponding generating Lie algebra $\mathfrak{g}_\alpha = \langle \mathcal{G}_\alpha \rangle_\mathrm{Lie} = \mathrm{span}(\mathcal{L}_\alpha)$ through the Lie closure. $\mathcal{L}_\alpha$ is the set created from calculating the nested commutators of $\mathcal{G}_\alpha$ and is a basis of $\mathfrak{g}_\alpha$.
    Any $\mathcal{G}_\alpha$ is a set of operators from $\mathcal{M}\mathfrak{su}(2^n)$. Subsequently, $\mathcal{L}_\alpha \subseteq \mathcal{M}\mathfrak{su}(2^n)$ and therefore $P^\alpha \subseteq \mathcal{M}\mathrm{SU}(2^n)$, meaning all terms in $F$ are invariant under the symmetry operations $\mathcal{M}$. Corresponding dynamical Lie groups are defined as 
    \begin{equation}
    \mathcal{P}_\alpha = \mathrm{e}^{\mathfrak{g}_\alpha} \equiv \left \{ \mathrm{e}^A, \quad A \in \mathfrak{g}_\alpha = \mathrm{span}(\mathcal{L}_\alpha) \right \}.
    \end{equation}
    The term
    \begin{equation}
        P^{\hat{\alpha}} = P^{(1,1,...,1)} = P_\mathrm{M} \cdots P_2 P_1,
    \end{equation}
    has $\mathcal{G}_{\hat{\alpha}} \equiv \mathcal{G}$ and consequently $\mathcal{L}_\alpha \subseteq \mathcal{L}_{\hat{\alpha}}$ for all $\alpha$. This means that $P^{\hat{\alpha}}$ is the most expressive term in $F$ and all possible operators composing $F$ are able to be generated in $P^{\hat{\alpha}}$. In other words all $\mathcal{P}_\alpha \in \mathcal{P}_{\hat{\alpha}}$.\\
    The next step is to further decompose the exponential $P^{\hat{\alpha}} = \mathrm{e}^Q$, where 
    \begin{equation}
        Q = \sum^{\mathrm{dim}(\mathcal{L}_{\hat{\alpha}})} d_{i}L_{i},
    \end{equation}
    with $L_{i} \in \mathcal{L}_{\hat{\alpha}}$ and $d_{i} \in \mathbb{R}$. Using a Taylor expansion we get
    \begin{equation}
        \mathrm{e}^Q = \sum^\infty_{k=0} \frac{1}{k!} Q^k.
    \end{equation}    
    Rewriting the expansion gives the finite series
    \begin{equation}
        \mathrm{e}^Q = \sum^D_{k=0} e_{k} B_{i}.
        \label{eq:singleB}
    \end{equation}
    The length $D$ is upper-bounded by $4^n-1$, which is the size of the full Lie algebra $\mathfrak{su}(2^n)$ and the coefficients $e_{k} \in \mathbb{C}$ are restricted so that $\mathrm{e}^Q$ is unitary. The operators $B_{i}$ make up the set $\mathcal{B}_{\hat{\alpha}}$. Since $Q$ is also a linear combination of matrices from $\mathcal{L}_{\hat{\alpha}}$, calculating the coefficients $e_{k}$ is not trivial, but the set $\mathcal{B}_{\hat{\alpha}}$ can be acquired by decomposing each $Q^k$ into linearly independent operators. For $Q^0$ we get the identity $\mathds{1}$ and $Q^1$ gives all elements of the set $\mathcal{L}_{\hat{\alpha}}$, which are part of $\mathcal{B}_{\hat{\alpha}}$. Higher order terms produce products $L_{i}L_{j}$. These products are included in $\mathcal{B}_{\hat{\alpha}}$ if they are linearly independent. These calculations are equivalent to calculating the associative closure of $\mathcal{L}_{\hat{\alpha}}$.

    By calculating the operators from the set $\mathcal{B}_{\hat{\alpha}}$, we have indeed calculated all possible $B_{i}$ that could turn up in any term of $F$, because $P^{\hat{\alpha}}$ is the most expressive term. Therefore, $\mathcal{B}_{\hat{\alpha}}$ is equivalent to the universal set $\mathcal{B}$. This means we can formulate the produced matrix function as
    \begin{equation}
        F = \sum^D_{k=0} f_{k} B_{i},
    \end{equation}
    with $B_{i} \in \mathcal{B}$, scalar coefficients $f_{i} \in \mathbb{C}$ and series length $D = \mathrm{dim}(\mathcal{B})$. While the coefficients $e_{i}$ in eq. \eqref{eq:singleB} are under the conditions of the unitarity of $\mathrm{e}^Q$ and thus not universal, the sum in eq. \eqref{eq:FsumP} has $2^M$ terms with assumably $N_\mathrm{p} = 3 \cdot M + 3$ independent scalar parameters. Given that $N_p \geq \mathrm{dim}(\mathcal{B})$, we can conjecture that the coefficients $f_{i}$ are universal and independent. We have intensely tested this for various inputs and no deviation has been found from the assumption that
    \begin{equation}
        F \in \mathrm{span}_\mathbb{C}(\mathcal{B}).
    \end{equation}
    It is also important to note for further consistency that 
    \begin{equation}
        \forall B \in \mathcal{B}, \forall S \in \mathcal{M}: SBS^{-1} = B
    \end{equation} 
    and $F$ is subsequently $\mathcal{M}$-invariant.\\
    
    We can now explain the observations from Table \ref{tab:GQSPres} by calculating the dimensions of each $\mathcal{B}$ using Algorithm \ref{alg:1}, which are listed in Table \ref{tab:Bdim}.  The sets $\mathcal{B}$ created by the different generating sets $\mathcal{G}$, from section \ref{sec:GenSets} can span the complete symmetric Lie algebra in the cases of $Z_\mathrm{2,xz}$ and $S_\mathrm{n}$ and the number of circuit parameters is consistent with the dimension of each $\mathcal{M}\mathfrak{su}(2^n)$, with the only addition being the identity matrix. In case of the symmetry $C_\mathrm{n}$, the generators cannot construct the space of $\mathcal{M}\mathfrak{su}(2^n)$, but all terms in the Hamiltonian are in $\mathcal{B}$, leading to a lower circuit depth than initially predicted.

    Since $\mathcal{B}\backslash \mathds{1} \nsubseteq \mathcal{M}\mathfrak{su}(2^n)$(products of $L_{i}$ are not generally anti-hermitian), $F$ is also not (anti-)hermitian. Hermiticity in the output can still be achieved through a hermitian circuit construction.
    \begin{algorithm}
        \SetKwInOut{Input}{Input}
        \SetKwInOut{Output}{Output}
        \caption{Circuit expressibility}\label{alg:1}
        \Input{Set of circuit generators $\mathcal{G}$, Matrix $M$}
        \Output{Basis set $\mathcal{B}$}
        $\mathcal{L} \leftarrow \mathcal{G}$\\
        \tcp{Step 1: Calculate commutative closure}
        \For{$L_{i} \in \mathcal{L}$}
        {
            \For{$L_{j} \in \mathcal{L}$}
            {
                $L_{k} = [L_{i},L_{j}]$\\
                \If{$L_{k} \notin \mathcal{L}$ \textbf{and} $L_{k} \neq 0$}
                {
                    $\mathcal{L} \leftarrow \mathcal{L} \cup \{L_{k} \}$
                }
            }
        }
        \tcp{Step 2: Calculate associative closure}
        $\mathcal{B} \leftarrow \mathcal{L}$\\
        \For{$B_{i} \in \mathcal{B}$}
        {
            \For{$B_{j} \in \mathcal{B}$}
            {
                $B_{k} = B_{i} \cdot B_{j}$\\
                \If{$B_{k} \notin \mathcal{B}$ \textbf{and} $B_{k} \neq 0$}
                {
                    $\mathcal{B} \leftarrow \mathcal{B} \cup \{B_{k} \}$
                }
            }
        } 
        \tcp{Step 3: expressibility}
        \eIf{$M \in \mathrm{span}(\mathcal{B})$}
        {
            \Return{True, $\mathcal{B}$}
        }
        {\Return{False, $\mathcal{B}$}}
    \end{algorithm}
Calculating the set $\mathcal{B}$ and checking if an input matrix is expressible with the circuit generators $\mathcal{G}$ can be done using Algorithm~\ref{alg:1}. For the generator sets in this paper it was sufficient to calculate the associative closure of $\mathcal{L}$ using only 0th, 1st and 2nd order terms from $\mathrm{e}^Q$, resulting in only first order products of $L_{i}$ appearing in $\mathcal{B}$.

Important to note is that this method does not necessarily need to be used for symmetric inputs only but is more general in the sense that it can give indications about the expressibility of not only a set $\mathcal{G}$ from a symmetric Lie algebra $\mathcal{M}\mathfrak{su}(2^n)$, but a freely chosen set $\mathcal{G}$. This can be seen in Table \ref{tab:Bdim} which compares the dimension of $\mathcal{M}\mathfrak{su}$ and $\mathcal{B}$ for the different cases from the main text. dim($\mathcal{M}\mathfrak{su}$) is always 1 smaller than dim($\mathcal{B}$) because $\mathcal{B}$ includes the identity. The generators used for the $Z_\mathrm{2,xz}$ and $S_\mathrm{n}$ invariant encoding of the corresponding $H$ construct a $\mathcal{B}$ that is a basis of $\mathcal{M}\mathfrak{su}$ since all elements of $\mathcal{B}$ are linearly independent and are invariant under the symmetries of the generating set. For the cyclic symmetry the dimension of $\mathcal{B}$ for $n \geq 4$ is more than 1 smaller, meaning the circuit cannot implement the full $\mathcal{M}\mathfrak{su}$ but still is able to encode the terms in the Hamiltonian. This allows to widen the scope of checking expressibility circumventing the need for finding the full $\mathcal{M}\mathfrak{su}$.

\begin{table}[ht]
        \centering
        \begin{tabular}{c|cc|cc|cc}
            & \multicolumn{2}{c|}{$Z_\mathrm{2,xz}$} & \multicolumn{2}{c|}{$C_\mathrm{n}$} & \multicolumn{2}{c}{$S_\mathrm{n}$}\\\hline
            n & dim($\mathcal{M}\mathfrak{su}$) & dim($\mathcal{B}$) & dim($\mathcal{M}\mathfrak{su}$) & dim($\mathcal{B}$)& dim($\mathcal{M}\mathfrak{su}$) & dim($\mathcal{B}$)\\\hline
            2 & 5  & 6  & 5   & 6  & 5 & 6    \\
            3 & 19 & 20 & 11  & 10 & 9 & 10   \\
            4 & 71 & 72 & 37  & 28 & 18 & 19  \\
            5 & 271 & 272 & 103 & 68 & 27 & 28\\
            6 & 1056 & & 356 & 226 & 43 & 44  \\
            7 & 4160 &  & 1172 &  & 59 &  60  \\
            8 &      &  &      &  & 84 & 85 \\
        \end{tabular}
        \caption{Dimensions of the symmetric Lie algebra $\mathcal{M}\mathfrak{su}$ and basis set $\mathcal{B}$ of $F$ for different input symmetries $\mathcal{M}$ and circuit generators $\mathcal{G}$. Computational restrictions running algorithm \ref{alg:1} are the cause for empty values.}
        \label{tab:Bdim}
    \end{table}

A method to find more efficient circuits is to recursively adapt the set of generators. First, a set of generators is needed that can implement the input matrix, for example the terms in the Hamiltonian such as done in the main text. From the calculated $\mathcal{B}$ terms are selected that have a lower gate count when implemented and the implementability is checked again. This can be continued until the gate count of the ansatz is satisfactory.

\subsection{Scaling}
    Two distinct scalings of the algorithm \ref{alg:1} and quantum circuit complexity have to be differentiated. Algorithm \ref{alg:1} is completely executed on classical hardware, which gives of course an exponential scaling in system size $n$ or alternatively linear in $N = 2^n$. Additionally, the runtime of the first part of the algorithm is directly dependent on the dimension of the DLA of $\mathcal{G}$. In total $\mathrm{dim}(\mathcal{L})^2$ commutators have to be checked. DLA-sizes are currently under intense research \cite{Wiersema_2024,allcock2024dynamicalliealgebrasquantum}. Under the assumption, that only $Q^k$ until $k=2$ is sufficient for producing the elements of $\mathcal{B}$, an additional $\mathrm{dim}(\mathcal{L})^2$ products have to be checked for linear independence. The algorithm scales as $\mathcal{O}(N\mathrm{dim}(\mathcal{L})^2)$, depending on the size of the DLA and validity of $k \leq 2$.

    On the other hand, quantum circuit complexity will scale as $\mathcal{O}(\mathrm{dim}(\mathcal{B})) = \mathcal{O}(\mathrm{dim}(\mathcal{L})^2)$. If the lower bound is reached and the generators are sufficiently occurring in the ansatz sequence, then
    \begin{equation}
        M_\mathrm{thres} = \left \lceil \frac{q \cdot \mathrm{dim}(\mathcal{B})}{3}-3 \right \rceil \, .
    \end{equation}
    For non-hermitian inputs $q=2$ and hermitian inputs $q=1$ since hermitian matrices have half the amount of degrees of freedom compared to a non-hermitian matrix in the general case.

    DLAs of Heisenberg Hamiltonians in chain configuration with (non)-periodic boundary conditions scale both $\mathcal{O}(4^n)$ \cite{Wiersema_2024,Mansky_2025}. Their exponential growth is also seen in Table \ref{tab:Bdim} and is transferred to the size of $\mathcal{B}$. In contrast, the permutation invariant circuit will only grow with a scaling as $\mathcal{O}(n^3)$ \cite{Mansky_2025}. Simpler generating sets, for example operators from Ising Hamiltonians often have polynomial scaling, also in low-symmetric configurations\cite{allcock2024dynamicalliealgebrasquantum}.

\subsection{Alternative approach to matrix function expressibility}
The method described in the appendix \ref{app:Bcalc} is useful to find out if a set of generators is in principle sufficient to encode an input matrix independent of the occurrences in the circuit. This can be a limiting assumption if the circuits are shallow or a generator is barely used in the ansatz sequence. In the following, a procedure is outlined to find the expressivity of a specific circuit generator sequence in the ansatz.
We start again with the variational matrix function
    \begin{equation}
        F = \prod_{i=1}^{M} \left ( a_{i} \mathds{1}^{\otimes n} + b_{i} P_{i} \right ),
        \label{eq:Ffunc}
    \end{equation}
but instead of first calculating the DLA and then decomposing the exponential, we decompose the exponentials first using
    \begin{equation}
        P_{i} = \mathrm{e}^{i \theta G_{i}} = \sum_{k=0}^{q_{i}} c_{i,k}T_{i,k} = c_{i,0} \mathds{1} + \sum_{k=1}^{q_{i}} c_{i,k}T_{i,k}
    \end{equation}
The $T_{i,k}$ are all elements from the set created by $\{G_{i}^k\}$ for $k=0$ to $k=q_{i}$ until no new linearly independent elements appear. Often it suffices to include only $\{\mathds{1},G,G^2\}$. Inserting the decomposition into eq. \eqref{eq:Ffunc} gives

    \begin{align}
        F &= \prod_{i=1}^{M} \left ( a_{i} \mathds{1}^{\otimes n} + b_{i} \sum_{k=0}^{q_{i}} c_{i,k}T_{i,k} \right )\\
          &= \prod_{i=1}^{M} \left ( \sum_{k=0}^{q_{i}} c_{i,k}T_{i,k} \right )\\
          &= \sum_\alpha d_\alpha T_\alpha\\
          &= \sum^{\mathrm{dim}(\mathcal{B})}f_{j} B_{j}
    \end{align}
    with $\alpha = (\alpha_1,\alpha_2,...,\alpha_\mathrm{M})$, $\alpha_{i} \in \{0,...,q_{i}\}$ and $T_\alpha = T_{0,\alpha_1}T_{1,\alpha_2} \cdots T_{M,\alpha_\mathrm{M}}$, which are in total $L = \prod q_{i}$ terms. Calculating all terms $T_\alpha$ and selecting a linearly independent set $\mathcal{B} = \{B_{i}\}$ results in a (restricted) basis of the matrix function $F$. Just as before the question is if the coefficients $f_{j}$ are universal and independent. While this method can construct a basis set $\mathcal{B}$ that is directly linked to the sequence of generators, it uses significantly more computational resources to calculate all $T_\alpha$.

\end{document}